\documentclass[a4paper,11pt]{article}
\pdfoutput=1 

\usepackage{jheppub} 
\usepackage[T1]{fontenc}
\usepackage[T1]{fontenc} 
\usepackage{tikz}
\usepackage{tikz-cd}

\newcommand{\cA}{\mathcal A}

\newcommand{\cC}{\mathcal C}

\newcommand{\cF}{\mathcal F}
\newcommand{\cG}{\mathcal G}

\newcommand{\cI}{\mathcal I}

\newcommand{\cL}{\mathcal L}

\newcommand{\cN}{\mathcal N}
\newcommand{\cO}{\mathcal O}

\newcommand{\cZ}{\mathcal Z}

\newcommand{\bea}{\begin{eqnarray}}
\newcommand{\eea}{\end{eqnarray}}

\usepackage[utf8]{inputenc}
\usepackage{hyperref}
\usepackage{amsfonts}
\usepackage{amsmath}
\usepackage{amssymb}
\usepackage{mathrsfs}  
\usepackage{color}
\usepackage{physics}
\usepackage{multicol}
\usepackage{tikz}
\newcommand{\ee}{\mathrm{e}}
\newcommand{\ii}{\mathrm{i}}

\title{\textbf{Unitary matrix models and random partitions: Universality and multi-criticality}}
\author{\textsc{Taro Kimura} and \textsc{Ali Zahabi}}

\affiliation{Institut de Math\'ematiques de Bourgogne, Universit\'e Bourgogne Franche-Comt\'e, France}

\abstract{
The generating functions for the gauge theory observables are often represented in terms of the unitary matrix integrals. In this work, the perturbative and non-perturbative aspects of the generic multi-critical unitary matrix models are studied by adopting the integrable operator formalism, and the multi-critical generalization of the Tracy--Widom distribution in the context of random partitions. We obtain the universal results for the multi-critical model in the weak and strong coupling phases. The free energy of the instanton sector in the weak coupling regime, and the genus expansion of the free energy in the strong coupling regime are explicitly computed and the universal multi-critical phase structure of the model is explored. Finally, we apply our results in concrete examples of supersymmetric indices of gauge theories in the large $N$ limit.
}
\begin{document} 
\maketitle

\flushbottom

\section{Introduction and summary}
The construction and counting of the gauge invariant observables, such as BPS operators in supersymmetric gauge theories, have been central problems in gauge theory and string theory, for decades. Often, the counting problems in gauge theory enjoy combinatorial features which makes the problems more tractable. For example, the counting of single-trace and multi-trace operators in supersymmetric gauge theories can be seen as the counting of letters and words in combinatorics, in the plethystic program \cite{feng2007counting,benvenuti2007counting}.

The plethystic program provides a framework to study the generating function of the gauge theories.
An essential step in this framework is applying the Weyl integration formula in the gauge theory \cite{skagerstam1984largen,sundborg2000hagedorn,Aharony:2003sx}, to obtain the full generating function of the multi-trace operators as a group/matrix integral of the plethystic exponentiation of the single-trace operator generating function.
The potential of this matrix model is turned up to be the double-trace potential,
and in the weak coupling limit, one can approximate the pairwise interaction potentials between the eigenvalues of the matrix model with a single-trace potential, i.e. the Gross--Witten--Wadia (GWW) model~\cite{Gross:1980he,Wadia:1980cp} and its generalization to higher order polynomial potentials. 

One important aspect of the counting problems is the asymptotic behavior of the generating functions and their multiplicities.
In the context of gauge theory, to study the thermodynamic aspects such as confinement/deconfinement phase transition and the perturbative/non-perturbative aspects one need to develop the asymptotic analysis for the gauge theory indices in the limit of large parameters and study the singularities of the generating functions. In the contexts of the AdS/CFT correspondence, the asymptotic aspects of the gauge theory contain interesting and important information for the gravity dual theories.

A recent interesting counting problem in gauge theory is the computation of the index of the $\cN=1$ four-dimensional superconformal field theories (SCFT), the generating function of the BPS operators which are annihilated by one supercharge.
In the context of AdS/CFT correspondence, the large $N$ and large charge limits of the superconformal indices, as a candidate of the microscopic explanations of the black holes entropy and phase transition have attracted a lot of interests and many different methods are developed toward the asymptotic study of the index, see \cite{zaffaroni2020ads} for a review.

In this work, we introduce an analytic approach, based on the machinery of the integrable operator formalism in random matrices and random partitions to study the universal features in the phase structure of gauge theories.
The techniques from random partitions and their asymptotics provide a natural framework for study of the dynamics of the gauge theory. Having obtained an alternative formulation of the generic unitary matrix model based on the Schur partition, i.e. a random partition obeying the Schur measure~\cite{Borodin:2000IEOT}, in this paper, we apply this machinery in the asymptotic analysis of the supersymmetric indices and their unitary matrix integrals, and explore the thermodynamics of the gauge theory and the phase structure. 
We study the finite and large $N$ asymptotics and associated phase structure of the unitary matrix models, mainly the generalized GWW model. Our results include computation of the partition function of the generic unitary matrix model which have wide applications in gauge theories. The generating functions for the supersymmetric indices of the gauge theory such as superconformal index are often represented in terms of the unitary matrix integrals with double trace potential. In the limit of weak interactions between the eigenvalues, they can be approximated by the matrix models with the single-trace potential, i.e. the generalized Gross--Witten--Wadia model. 
We aim to shed light on the implications of the universality for the phase structure of the gauge theory. 
In fact, one can imagine that the fluctuations of the random partitions in the bulk or at the edges explain a possible phase transition in the associated gauge theories. The generating function of the indices and their matrix integrals in the large $N$ limit can be represented in terms of the Fredholm determinants. The asymptotic analysis of the Fredholm determinants and emergence of the Tracy--Widom (TW) distribution explain the phase structure of the gauge theories. 

Precisely speaking, the edge and the bulk fluctuations in the Schur partitions, described by the Fredholm determinants with sine, and Airy kernels, respectively, can be applied in the study of critical dynamics of a class of gauge theories with a generic single-trace unitary matrix integrals near its critical point. Moreover, the finer phase structure of the matrix models known as the multi-critical generalization emerges from the smaller scale fluctuations and they are studied using the higher Airy kernels and their asymptotics.
In fact, the critical dynamics of the matrix model is encoded in the asymptotic behavior, i.e. the right and left tails of the TW distribution. In the large $N$ limit, the sharp phase transition is implicit in the different behavior of the tails of the distribution. This phase transition is replaced by a smooth cross-over, in the intermediate domain of the distribution at finite $N$. Moreover, the perturbative and non-perturbative aspects of the gauge theory are obtained from the finite $N$ corrections to the asymptotic behavior of the two tails of the distribution. More precisely, the genus expansion of the free energy is obtained via the perturbative $1/N$ corrections in the asymptotic analysis of the left tail
of the distribution.
The exponential order corrections and related instanton effects are obtained from the right tail of the distribution, using the Airy function approximation.

In concrete example of the GWW model and its generalization, our result is indicating a universal third-order phase transition, and we compare it with the model-dependent results in the literature which are based on different plausible methods such as the Coulomb gas method and saddle-point analysis of the matrix models, and we find a good agreement, when expanding the results around the critical point.
As an application of our result, we study the fine structure of the Hagedorn phase transition in two concrete examples of the gauge theory indices. We find that the phase structure of the generalized GWW model captures the leading order singularities of the Hagedorn phase transition. Based on this observation, we compute the free energy of these gauge theories in different regimes.
In the finer double-scaling regimes, similar results for the multi-critical dynamics are obtained.
\subsection*{Summary}
Let us briefly summarize the main results of this paper. We will collect the main ideas, tools and results which are roughly expressed, and refer to the bulk of the paper for the exact expressions.
By using the character expansion formula, we can write the matrix integral with multi-critical potential in terms of the sum over partitions with the Schur measure. In fact, imposing the constraint on the largest entry of the partition $\lambda_1$, the summation over the random partition is rewritten, modulo a normalization factor, as the unitary matrix integral~\cite{Borodin:2000IEOT},
\begin{align}
  \mathcal{Z}_N 
  \propto \sum_{\substack{\lambda \in \mathscr{Y}\\ \lambda_1 \le N}} s_\lambda(\mathsf{X}) s_\lambda(\mathsf{X})
  = \int_{\mathrm{U}(N)} \hspace{-1em} \dd{U} \exp\qty( \sum_{n=1}^\infty \frac{1}{n}f(q_1^n, q_2^n, ...)\qty( \tr U^n + \tr U^{-n}) ),
  \label{pf}
\end{align}
for definitions see Section \ref{sec:Schur random partition and generalized GWW model}. An unrefined version of the partition function \eqref{pf} is recently studied in \cite{santilli2020exact}.
Defining the free energy by $\mathcal{F}=\lim_{N \to \infty} N^{-2} \log  \mathcal{Z}_N$, we show that
the right/left ($\pm$) edge fluctuation contributes to the free energy, up to a scaling factor $N^{-2}$, as
\begin{align}
    \mathcal{F}\sim\lim_{s \to \pm\infty} \log F_p(s),
    \qquad \text{where} \qquad
    s= \frac{ (\beta_c-\beta) N}{(\alpha_p N)^{\frac{1}{p+1}}},
\end{align}
and $F_p$ is the higher-order Tracy--Widom distribution \cite{Periwal:1990qb,Claeys:2009CPAM,LeDoussal:2018dls,cafasso2019fredholm}, and $\alpha_p(q_1, q_2, ...)$, $\beta(q_1, q_2, ...)$ are some model-dependent parameters and are explicitly expressed in terms of the couplings~\cite{kimura2020universal}.
The main result of our study is that the matrix model \eqref{pf}, undergoes a multi-critical phase transition 
at the critical point $\beta=\beta_c$. This phase transitions can be explained from the asymptotic behavior of the higher-order Tracy--Widom distribution, 
\begin{equation}
\label{}
F_p(s) \sim
  \begin{cases}
   1-\mathcal{O}\left(s^{-\frac{p+1}{p}}e^{-s^{\frac{p+1}{p}}}\right) & \quad s \rightarrow +\infty \\
   \\
   \mathcal{O}\left(e^{-|s|^{\frac{2(p+1)}{p}}}\right) & \quad s \rightarrow -\infty \\
  \end{cases}.
\end{equation}

The above leading asymptotic behavior of the higher-order TW distribution and the following multi-critical double-scaling parameter,
$
s=\alpha_p^{-\frac{1}{1+p}} (\beta_c- \beta)\ N^{\frac{p}{p+1}},
$
imply that the 
free energy in two regimes $s\to +\infty$ ($\beta<\beta_c$), and $s\to -\infty$ ($\beta>\beta_c$), modulo an additive constant, is given by 
\bea
    \cF \sim \begin{cases}
    \displaystyle
        \cO(e^{-cN}) &  \beta<\beta_c\\[1em] \displaystyle
      \alpha_p^{-2/p}|\beta_c-\beta|^{2(p+1)/p}+\cO(N^{-2}) &  \beta>\beta_c
        \end{cases}.
\eea
The above free energy implies a discontinuity in the $(2(p+1)/p)$-th derivative of the free energy at $\beta=\beta_c$ and a multi-critical phase transition of the order $(2(p+1)/p)$ at this critical point. One can possibly make a mathematical sense of the fractional derivative in the fractional calculus. Alternatively, the order of the multi-critical phase transition, can be replaced by $\lfloor\frac{2(p+1)}{p}\rfloor = 3$ for any finite $p>2$. At the classic case of $p=2$, a universal third-order phase transition is emerged from the fluctuations at scale $N^{-2/3}$. This can be seen as a universal generalization of the GWW model~\cite{Gross:1980he,Wadia:1980cp}, see also \cite{zahabi2016new}. For higher $p>2$, the fluctuations at scale $N^{-\frac{p}{p+1}}$ creates a multi-critical phase transition and as we increase $p$, the order of the phase transition monotonically decreases. At infinite $p$, the fluctuations are of order $N^{-1}$ and they causes a second order phase transition. We focus on the two fixed-points of $p=2$ and $p=\infty$ and their possible physical implications in the general setting and in concrete examples.




The rest of the paper is organized as follows. In Section~\ref{sec:index_matrix}, the matrix integral representation of the supersymmetric indices is explained. In Section~\ref{sec:op_form}, the Schur partition and its asymptotics are reviewed. In Sections~\ref{sec:crit_dyn} and \ref{multi section} the critical and the multi-critical dynamics of the generalized GWW model are discussed. In Section~\ref{example}, we study the asymptotics of the double-trace matrix model in some concrete examples and explicit results about the free energy and phase structure are obtained.

\section{Gauge theory indices and matrix models}\label{sec:index_matrix}
This Section is a brief review about the two classes of unitary matrix models, namely the generalized double-trace and generalized GWW model and their relations. These two classes are related to the partition functions of the gauge theories on compact manifolds and indices of the supersymmetric gauge theories.
\subsection{Counting observables in gauge theory}
The observables of the gauge theories are the set of gauge invariant operators and their correlation functions. We consider four-dimensional gauge theories and their observables. These are formed of single trace of the products of operators, called single-trace operators and several single-trace operators multiplied to each other, called multi-trace operators.
The grand-canonical partition function of the multi-trace operators is obtained from the plethystic exponentiation of the generating function of the single-trace operators $f(q_i)$, where $q_i$ is the short notation for $\{q_i\}$, the collection of fugacity factors~\cite{benvenuti2007counting,feng2007counting}, and then integrating over the Haar measure of the gauge group to project onto the gauge-singlets~\cite{skagerstam1984largen,sundborg2000hagedorn,Aharony:2003sx}, known as Weyl integration formula,
\bea
\label{}
\mathcal{I}(q_i) = \int_{G}  \dd{g}\ \exp\left(\sum_{n=0}^\infty \frac{1}{n}f(q_i^n) \chi_{R}(g^n)\right),
\eea
where $g\in G$, and in our study $G$ is to be considered as the unitary gauge group $\mathrm{U}(N)$ or their product, and $\chi_{R}(g)$ is the character in $R$-representation. In Section \ref{example}, we discuss some concrete examples in four-dimensional $\mathcal{N}=1$ SCFT.

\subsection{Matrix integral representation}
In this Section, the counting problem of the BPS operators and their generating functions in terms of matrix integral are reviewed.
We consider the group integral that appears as the partition function of $G=\mathrm{U}(N)$ gauge theories with the adjoint representation matter, and by using the character formula in the adjoint representation in terms of the characters in the fundamental representation $\chi_{\text{adj}}(g)= \chi_{F}(g)\chi_{F}^*(g)$, the group integral recasts to a matrix integral, 
\bea
\label{index PE}
\mathcal{I}(q_i) = \int_{\mathrm{U}(N)} \hspace{-1em} \dd{U}\ \operatorname{PE} \left[f(q_i;U)\right],
\eea
where by using the Weyl parameterization, the integral over the Haar measure can be written as an integral over the eigenvalues $\theta_i$ in the diagonal matrix $U= \operatorname{diag}\left(e^{\mathrm{i} \theta_i}\right)_{i=1}^N$,
\bea
\int_{\mathrm{U}(N)} \hspace{-1em} \dd{U} = \frac{1}{N!}\int \prod_{i=1}^N \frac{d\theta_i}{2\pi}\prod_{1\leq i<j\leq N} |e^{\mathrm{i} \theta_i}- e^{\mathrm{i} \theta_j}|^2,
\eea
and the plethystic exponential is defined by
\bea
\operatorname{PE} [f(q_i;U)]= \exp\left[\sum_{n=1}^\infty \frac{1}{n} f (q_i^n) (\tr U^n\tr U^{-n}) 
\right].
\label{PE def}
\eea
The above matrix integral appears as the generating function for the BPS multi-trace operators in free gauge theories. In particular, we are going to consider, in Section \ref{example}, two explicit classes of examples of this, which are i) BPS operators in chiral ring of the free four-dimensional $\cN=1$ SCFT and ii) the BPS operators that are annihilated by one supercharge in $\cN=1$ SCFT, in particular the sixteenth BPS operators in $\cN=4$ supersymmetric Yang--Mills (SYM) theory.  

Let us define a closely related model, the generalized GWW matrix model, 
\bea
\label{index gGWW}
\mathcal{Z}(q_i) = \int_{\mathrm{U}(N)} \hspace{-1em} \dd{U} \exp\left[N\sum_{n=1}^\infty t_n(q_i) \qty( \tr U^n+ \tr U^{-n} )\right].
\eea
Notice that, in general, the $\cO(1)$ couplings $t_1, t_2, ...$ can be any univariate or multivariate function of the parameters of the model. 
In the rest of this paper, we absorb the $N$ dependence into the definitions of the couplings and thus drop the overall factor $N$. Truncating to the case $n=1$, $\mathcal{Z}(q_i)$ reduces to a effective theory of GWW model,
\bea
\label{}
\mathcal{Z}(q_i) \sim \cZ_{\text{GWW}}= \int_{\mathrm{U}(N)} \hspace{-1em} \dd{U} \exp \left[ t_1 (q_i)(\tr U+ \tr U^{-1})\right].
\eea

The gauge theory indices as the double-trace matrix integral \eqref{index PE} is related to the generalized GWW matrix integral, either effectively at weak coupling in the large $N$ limit or exactly by the Hubbard-Stratonovich transformation~\cite{liu2004fine,alvarez2006black}. In this study, we focus on the former approximation and leave the latter relation for the future studies. The double-trace matrix model and the generalized GWW model share equilibrium conditions. In fact, 
in the large $N$ limit, $\mathcal{I}(q_i)$ and $\mathcal{Z}(q_i)$ are equivalent, upon identification
\bea
t_n(q_i) =\frac{f(q^n_i)}{n} \langle\tr U^n\rangle,
\eea
and they have a similar phase structure~\cite{Aharony:2003sx}. Moreover, the truncated case $n=1$ of the double matrix integral \eqref{index PE}, and the GWW model,  are dominant in the asymptotic analysis of the generalized models with higher degrees, $n>1$, see \cite{Aharony:2003sx} for more details. Being aware of exact relation between double trace matrix models and GWW models by 
the Hubbard--Stratonovich transformation and saddle point approximation~\cite{alvarez2005finite}, in this article we are going to focus on the generalized GWW matrix model and its asymptotic analysis and interpreting the results as the effective theory for the double trace models at the weak coupling limit. We hope to return to this problem in future and precisely apply the results of generalized GWW model to the double-trace model, using the transformation.


Having explained the connection between matrix integral representation of gauge indices and the generalized GWW model, in the rest of the paper, we are going to explore the generalized GWW model and its possible implications for gauge theory indices. We introduce random partitions with Schur measure to study the phase structure of the generalized GWW model.  Our goal is to study the large $N$ limit of the superconformal index and its phase structure, using the asymptotic analysis of matrix integral. In the next Section, we use the asymptotic analysis is performed using the methods of random partitions and integrable operator formalism. This approach uncovers the universal features of the phase structure.

\section{Integrable operator formalism in matrix models}\label{sec:op_form}
In this part, we review, without details of the proofs, the results obtained in recent works~\cite{kimura2020universal,Betea:2020} about the random partitions and its asymptotics, and then adopt them to study the phase structure of gauge theories. The first step in the random partition realization of the gauge theory is to think of the random partition as discretization of the matrix integral representation of the partition function of the gauge theory. More precise relation between the matrix integral and the random partition is via the character expansion formula.

\subsection{Schur measure random partition and generalized GWW model} \label{sec:Schur random partition and generalized GWW model}
Let $\mathscr{Y}$ be a set of partitions, and $\mathsf{X} = (\mathsf{x}_i)_{i \in \mathbb{N}}$ be the set of the parameters and the Miwa variables defined by
\begin{align}
    t_n = \frac{1}{n} \sum_{i=1}^\infty \mathsf{x}_i^n.
    \label{eq:Miwa_var}
\end{align}
Then, we define our main object, the partition function of the constrained partitions, via the Schur function $s_\lambda$, as
\begin{align}
  \mathcal{Z}_N 
  = \mathbb{P} (\lambda_1<N)= Z^{-1} \sum_{\substack{\lambda \in \mathscr{Y}\\ \lambda_1\leq N}} s_\lambda(\mathsf{X}) s_\lambda(\mathsf{X}),
  \label{pf prob}
\end{align}
where $Z$ is the normalization of the Schur measure random partition and it is given by,
\begin{align}
    Z 
    = \sum_{\lambda \in \mathscr{Y}} s_\lambda(\mathsf{X}) s_\lambda(\mathsf{X}) 
    = \prod_{1 \le i, j \le \infty} \qty( 1 - \mathsf{x}_i \mathsf{x}_j )^{-1} 
    = \exp \qty( \sum_{n = 1}^\infty n \, t_n^2 )
    \, .
    \label{eq:Schur_part_fn}
\end{align}
We will elaborate on the physical interpretation of the partition functions $Z$ and $\cZ$ in  Section~\ref{Asymptotic analysis of partitions and dynamics of gauge theory}.
Using the character expansion, the Schur partition can be written in terms of the unitary matrix integral,
\bea
\cZ_N = Z^{-1}\int_{\mathrm{U}(N)} \hspace{-1em} \dd{U} \exp\qty( \sum_{n=1}^\infty t_n\qty( \tr U^n + \tr U^{-n}) ).
\label{}
\eea

Let us define the Fredholm determinant on a specific domain $I \subset \mathbb{R}$,
\begin{align}
    \det( 1 - K)_I = \sum_{n = 0}^\infty \frac{(-1)^n}{n!} \int_{I^{n}} \prod_{i=1}^n \dd{x_i} \det_{1 \le i, j \le n} K(x_i, x_j)
    \, ,
    \label{eq:Fredholm_det_def}
\end{align}
where $K$ is the integral operator with the kernel $K(x_i,y_j)$. Then, the Schur partition function can be written
as a discrete analog of the original Fredholm determinant~\eqref{eq:Fredholm_det_def}~\cite{Borodin:2000IEOT}, 
\bea
\cZ_N 
= \det\left(1- K\right)_{I},
\label{two-fold}
\eea
where the domain $I$ is defined as $I=(N + \frac{1}{2},\infty)$, and $K(k,l)$ is given by \cite{Okounkov:2001SM},
\begin{subequations}\label{eq:Schur_kernel}
\begin{align}
    K(k,l) & = 
    \frac{1}{(2 \pi \ii)^2}
    \oint_{|z|>|w|} \hspace{-1.5em} \dd{z} \dd{w} \,
    \frac{\mathsf{K}(z,w)}{z^{k + 1/2} w^{- l + 1/2}}
    \qquad \qty(k, l \in \mathbb{Z} + \frac{1}{2})
    \, , \\
    \mathsf{K}(z,w) & = \frac{\mathsf{J}(z)}{\mathsf{J}(w)} \, \frac{1}{z - w}
    \, , \qquad
    \mathsf{J}(z) = \prod_{n = 1}^\infty \frac{1 - \mathsf{x}_n z}{1 - \mathsf{x}_n / z }
    \, .
\end{align}
\end{subequations}
The $\mathsf{J}(z)$ is called the wave function and it has the following mode expansion,
\begin{align}
    \mathsf{J}(z) 
    = \exp \left[ \sum_{n = 1}^\infty t_n\qty( z^n -  z^{-n}) \right]
    = \sum_{x \in \mathbb{Z}} J(x) \, z^{x}
    \, , \qquad
    J(x) =  \oint \frac{\dd{z}}{2 \pi \ii} \frac{\mathsf{J}(z)}{z^{x+1}}
    \, .
    \label{eq:J_exp}
\end{align}

\subsection*{Plancherel measure random partition and GWW model}
A particular example of the Schur measure with a single parameter
\bea
t_n = \mathfrak{q}\ \delta_{n,1},
\eea
is called the Plancherel measure with the following partition function,
\begin{align}
    \mathcal{Z}_N  = Z^{-1} \,\sum_{\substack{\lambda \in \mathscr{Y}\\ \lambda_1\leq N}} \mathfrak{q}^{2|\lambda|} \qty( \frac{\dim \lambda}{|\lambda|} )^2,
\end{align}
where  $\dim \lambda$ is the dimension of the irreducible representation parameterized by $\lambda$ of the symmetric group $\mathfrak{S}_\infty$, and
$|\lambda| = \sum_{i = 1}^\infty \lambda_i$, and
$Z = \ee^{\mathfrak{q}^2}$.

The Plancherel partition function with the constraint $\lambda_1 \le N$ has the matrix integral and Fredholm determinant representation,
\bea
\cZ_N= Z^{-1}\int_{\mathrm{U}(N)} \hspace{-1em} \dd{U} \exp\left(\mathfrak{q}  \qty( \tr U + \tr U^{-1}) \right)
=\det\left(1- K_\text{dB}\right)_{(N + \frac{1}{2},\infty)},
\eea
where the discrete Bessel kernel is obtained in \cite{Brodin:2000},
\begin{align}
    K_\text{dB}(k,l) = \mathfrak{q} \frac{J_{k - 1/2}(2\mathfrak{q})\, J_{l + 1/2}(2\mathfrak{q}) - J_{k+1/2}(2\mathfrak{q})\, J_{l - 1/2}(2\mathfrak{q})}{k-l}
    \, .
    \label{eq: dis Bess kern}
\end{align}
The wave function is then given by the Bessel function $J(x) = J_x(2 \mathfrak{q})$, with the generating function
\begin{align}
    \mathsf{J}(z) 
    = \ee^{\mathfrak{q} (z - z^{-1})}
    = \sum_{x \in \mathbb{Z}} J_x (2 \mathfrak{q}) \, z^x
    \, .
\end{align}
\subsection*{Difference equation for wave functions}
We observe that the wave functions of the Schur partitions satisfy the following differential and difference equations:
\begin{align}\label{eq:ODE_J}
    \qty[
    \sum_{n = 1}^\infty n\, t_n \qty( z^n + z^{-n}) - z \pdv{}{z}
    ] \mathsf{J}(z) & = 0,
    \qquad
    \qty[
    \sum_{n = 1}^\infty n\, t_n  \qty( \nabla_x^n + \nabla_x^{-n} ) - x
    ] J(x) = 0,
\end{align}
where we define the shift operator $\nabla_x f(x) = f(x+1)$ with $\nabla_x = \exp \qty( \partial_x )$.
In the case of the Plancherel measure, the wave function obeys the difference equation, a special case of~\eqref{eq:ODE_J},
\begin{align}
    \qty[ \nabla_x + \nabla_x^{-1} - \frac{x}{\mathfrak{q}} ] J_x(2\mathfrak{q}) = 0
    \, .
    \label{eq:ODE_B}
\end{align}

We remark that, as mentioned in \cite{kimura2020universal}, the differential/difference equation~\eqref{eq:ODE_J} is interpreted as a quantization of the spectral curve for the generalized GWW model,
\begin{subequations}\label{eq:sp_curve}
\begin{align}
    \Sigma_\text{GWW} 
    = \left\{ (x,z) \in \mathbb{C} \times \mathbb{C}^\times \mid \mathcal{H}(x,z) = 0 \right\}
\end{align}
with
\begin{align}
    \mathcal{H}(x,z) = \sum_{n = 1}^\infty n\, t_n  \qty( z^n + z^{-n} ) - x
    \, .
\end{align}
\end{subequations}

\subsection{Asymptotic analysis of partitions and dynamics of gauge theory}\label{Asymptotic analysis of partitions and dynamics of gauge theory}
In this Section, dynamics of gauge theory is obtained from the asymptotic analysis
of the Schur partition. 
The free energy in the large $N$ limit of the $\mathrm{U}(N)$ matrix model is defined by
\bea
\mathcal{F}=\lim_{N \to \infty} N^{-2} \log  \mathcal{Z}_N.
\label{F def}
\eea
To explore the phase structure of the gauge theory, we compute the free energy in different regimes of the parameter space, obtained from the asymptotic analysis of the random partitions. In term of the random partition, there are two contributions to the free energy, first a contribution from the continuum limit/limit shape of the random partition, obtained from the normalization factor of the matrix integral. Secondly, there is a contribution from the fluctuation around the limit shape. Depending on the region in the partition, the fluctuation behaves differently; in the bulk of the partition, the contribution is given by the Fredholm determinant of the sine kernel, and in the edge of the partition, it is the Fredholm determinant with Airy kernel, i.e. TW distribution. Thus, formally speaking, we have 
\bea
\cF= \cF_c+ \cF_f, 
\label{tot F}
\eea
where $\cF_c$ and $\cF_f$ denote the continuum and fluctuation free energy, respectively.

The continuum free energy is a global model dependent contribution
and can be computed using the matrix integral, for example in the generalized GWW model one can obtain from Eq.\eqref{eq:Schur_part_fn},
\bea
\cF_c= \log Z= \sum_{n=1}^\infty n\, t_n^2,
\label{cont F Schur}
\eea
notice that the $N$ dependence is implicit in the coupling $t_n$.
The fluctuation has different behavior in the bulk and edge of the partition, thus, we consider each case separately.
\subsubsection*{Bulk scaling limit and fluctuation}
\label{bulk fluc}
Now, let us consider the Plancherel measure random partition and discuss the scaling limit of the Bessel function. Let us define the discreteness parameter $\epsilon\sim N^{-1}$, to re-scale the parameters of the random partition, such that the large $N$ limit of the unitary matrix model corresponds to the continuum limit, $\epsilon \to 0$, of the random partition. Moreover, as we will explicitly explain later, the parameter $s$ in the domain of the Fredholm kernel \eqref{two-fold}, is in fact the double-scaling parameter and as $N$ tends to infinity, it tends to plus infinity in the bulk and plus/minus infinity at the right/left sides of the edge of the random partition. 

Using the scaling $(x,\mathfrak{q}) \to (x,\mathfrak{q}/\epsilon)$, the scaling limit of the difference equation~\eqref{eq:ODE_B} becomes
\begin{align}
    \qty[ \nabla_x + \nabla_x^{-1} + \cO(\epsilon) ] J_n\qty(2\mathfrak{q}/\epsilon) = 0    
    \, ,
\end{align}
which has the plane wave solutions
\begin{align}
    J_x(2\mathfrak{q}/\epsilon) \ \xrightarrow{\epsilon \to 0} \ 
    \begin{cases}
    \displaystyle
    \cos\qty(\frac{\pi}{2} x) = (-1)^{x/2} & (x \in 2 \mathbb{Z}) \\[1em]
    \displaystyle
    \sin\qty(\frac{\pi}{2} x) = (-1)^{x/2-1/2} & (x \in 2 \mathbb{Z} + 1) 
    \end{cases}.
\end{align}
Then, the scaling limit of the discrete Bessel kernel \eqref{eq: dis Bess kern} becomes the sine kernel
\begin{align}
    K(k,l) \ \xrightarrow{\epsilon \to 0} \ \frac{\sin \pi (k - l)/2}{\pi(k - l)/2}
    \, ,
\end{align}
where the normalization is fixed to be $K(r,r) = 1$.
Similar studies for the bulk scaling limit of the Schur measure random partition is performed in \cite{okounkov2003correlation}.

In the bulk, the dynamics of the gauge theory is governed by the sine random process and the fluctuation contribution to the free energy is given by the Fredholm determinant with the sine kernel, up to a scaling factor $N^{-2}$, as
\bea
\cF_f\sim\lim_{s\to\infty}\log\det \left(1-\hat K_{\text{sine}}\right)_{(s,\infty)},\qquad \text{where} \qquad \hat K_{\text{sine}}(x,y)=\frac{\sin\pi(x-y)}{\pi(x-y)}\nonumber\\.
\eea
The asymptotics of the Fredholm determinant as $s\to\infty$ is obtained in \cite{deift1997riemann,widom1994asymptotics},
\bea
\det \left(1-\hat K_{\text{sine}}\right)_{(s,\infty)}= 2^{1/12}\, e^{3\,\zeta'(-1)}\, s^{-\frac{1}{4}}\exp\left(-\frac{s^2}{2}\right)\left(1+\cO(s^{-1})\right).
\eea
This asymptotic result leads to the computation of the free energy in the bulk, the scaling parameter $s=\gamma N$, as
\bea
\cF&=& \cF_c+ \cF_f\nonumber\\
&=& \cF_c+\lim_{N\to +\infty}N^{-2}\log\det \left(1-\hat K_{\text{sine}}\right)_{(\gamma N,\infty)}\nonumber\\
&=&\cF_c - \frac{\gamma^2}{2}- \frac{1}{N^2}\left(-\frac{1}{4}\log N -\frac{\log \gamma}{4}  + \frac{\log 2 }{12} + 3\, \zeta'(-1)\right),
\eea
where $\cF_c$ is given by the normalization of the Schur measure random partition in Eq.\eqref{cont F Schur}.
Thus, we observe that in the large $N$ limit, the leading contribution of the bulk fluctuation in the free energy is non-zero. We will come back to this observation and discuss a possible interpretation of that, at the end of Section~\ref{infinite limit}.
\subsubsection*{Edge scaling limit and fluctuation}
Next, we discuss the scaling limit of the difference equations of the wave function and the kernel of the Fredholm determinant. We first consider the Schur partition. Let us use the parameter $\epsilon$ to re-scale the parameters and expand the shift operator,
\bea
(x,t_n) \to (x/\epsilon,t_n/\epsilon),\qquad \nabla_{x/\epsilon}^{\pm n} = \sum_{k = 0}^\infty \frac{(\pm n \epsilon)^k}{k!} \dv[k]{}{x}.
\eea
Then, the difference equation \eqref{eq:ODE_J} becomes
\begin{align}\label{diff eq}
    \qty[
    \sum_{k = 1}^\infty \alpha_{k} \, \epsilon^{k} \dv[k]{}{x} - (x - \beta)
    ] J\qty(\frac{x}{\epsilon}) = 0
    \, ,
\end{align}
where the coefficients are defined as
\begin{align}
    \alpha_k = \sum_{n=1}^\infty \frac{2n^{k+1}t_n}{k!}
    \, , \qquad
    \beta = \alpha_0 = \sum_{n=1}^\infty 2n \, t_n
    \, .
    \label{eq:alpha_beta}
\end{align}
In the scaling limit $\epsilon \to 0$, and by keeping $\alpha_{p'} = 0$ for $p' < p$, the difference equation \eqref{diff eq} becomes the following differential equation,
\begin{align}
    \qty[ \dv[p]{}{\xi} - \xi ] J\qty(\frac{\beta}{\epsilon} + \qty( \frac{\alpha_{p}}{\epsilon})^{\frac{1}{p+1}} \xi ) = 0
    \, ,
\end{align}
where $\xi= \alpha_p^{-\frac{1}{p+1}} \epsilon^{-\frac{p}{p+1}}(x -\beta)$ for $p \ge 2$.
Thus, we observe that the scaling limit of the wave function is given by the $p$-Airy function,
\begin{align}
    J\qty(\frac{\beta}{\epsilon} + \qty( \frac{\alpha_{p}}{\epsilon})^{\frac{1}{p+1}} \xi )
    \ \xrightarrow{\epsilon \to 0} \
\text{Ai}_{p}(\xi)= (-1)^{\frac{p}{2}+1}\int_{\cC} \frac{\dd\lambda}{2\pi \mathrm{i}}\, \exp{(-1)^{\frac{p}{2}}\frac{\lambda^{p+1}}{p+1}+\xi\lambda},
    \label{eq:p-Airy_lim}
\end{align}
where $\cC$ is an integral contour providing a convergent integral. The case $p = 2$ corresponds to the standard Airy function. The $p$-Airy function, $\text{Ai}_{p}\left((-1)^{\frac{p}{2}+1}\xi\right)$, satisfies the generalized Airy equation,
\bea
\dv[p]{\xi} \text{Ai}_{p}(\xi)= \xi \text{Ai}_{p}(\xi).
\eea
In fact, we can analyze the semi-classical behavior of the $p$-Airy function based on the reduced spectral curve,
\begin{align}
    \Sigma_{p\text{-Airy}} = \left\{ (x,y) \in \mathbb{C} \times \mathbb{C} \mid y^p - x = 0 \right\}
    \, .
    \label{eq:sp_curve_Airy}
\end{align}
This is obtained from the spectral curve of the GWW model~\eqref{eq:sp_curve} in the scaling limit discussed above, i.e., parametrize $w = \exp\qty(\epsilon y)$, then take the limit $\epsilon \to 0$ with tuning the parameters.
We remark that the $p$-Airy spectral curve~\eqref{eq:sp_curve_Airy} agrees with $(A_{p-1},A_0)$-type Argyres-Douglas theory~\cite{Cordova:2015nma,Wang:2015mra}.

Next, we discuss the scaling limit of the kernel.
Using the Eq.\eqref{eq:Schur_kernel}, we obtain the kernel in terms of the wave functions, 
\begin{align}
     \mathsf{K}(z,w) = \sum_{n,m \in \mathbb{Z}} \sum_{i = 1}^\infty J(n) J(m) \, z^{n - i } w^{- m + i - 1}, \quad K(k,l) = \sum_{i = 1}^\infty J\qty(k + i - \frac{1}{2}) J\qty(l + i - \frac{1}{2})
    \, ,
\end{align}
and then using the scaling limit of the wave function, one can obtain the scaling limit of the kernel as
\begin{align}
    K\qty( \frac{\beta}{\epsilon} + \qty( \frac{\alpha_{p}}{\epsilon})^{\frac{1}{p+1}} x,\, \frac{\beta}{\epsilon} + \qty( \frac{\alpha_{p}}{\epsilon})^{\frac{1}{p+1}} y)
    \xrightarrow{\epsilon \to 0} \
    \qty( \frac{\alpha_{p}}{\epsilon})^{\frac{1}{p+1}}
    K_{p\text{-Airy}}(x,y),
\end{align}
where the $p$-Airy kernel is defined as
\begin{subequations}
\begin{align}
    K_{p\text{-Airy}}(x,y) 
    & = \int_0^\infty \dd{z} \operatorname{Ai}_p(x+z) \operatorname{Ai}_p(y+z)
    \\
    & = \frac{1}{x - y} \sum_{q=0}^{p-1} (-1)^{q} \operatorname{Ai}_p^{(q)}(x) \operatorname{Ai}_p^{(p-q-1)}(y)
    \, ,
    \label{eq:Airy_kernel2}
\end{align}
\end{subequations}
and $\operatorname{Ai}_p^{(r)}(x)$ is the $r$-th derivative of the Airy function.
\subsection{Higher-order Tracy--Widom distribution}
The natural origin of the Airy kernel is in the theory of random matrices, in which the gap probability of the Airy process is given by the Fredholm determinant with the Airy kernel, 
\begin{align}
    F_2(s) = \det( 1 - K_\text{Airy})_{(s,\infty)}
    \, .
\end{align}
The probability distribution function of the largest eigenvalue and its statistical behavior in the scaling limit is given by the Tracy--Widom distribution~\cite{Tracy:1992rf}. 

Random partition, as a discrete analog of the random matrix~\cite{Borodin:2000IEOT}, has similar edge scaling behavior which is reflected in terms of the probability distribution of the largest entry (the first row) of the partition. In the scaling limit, the probability distribution of the largest entry of the Schur partition is obtained in \cite{kimura2020universal},
\begin{align}
    \lim_{\epsilon \to 0} \mathbb{P}\qty[ \frac{\lambda_1 - \beta / \epsilon}{(\alpha_p / \epsilon)^{\frac{1}{p+1}}} < s] 
    = \det( 1 - K_{p\text{-Airy}})_{(s,\infty)}
    =: F_p(s)
    \, ,
    \label{higher TW }
\end{align}
where $F_p(s)$ is a higher-order analog of the Tracy--Widom distribution~\cite{Periwal:1990qb,Claeys:2009CPAM,LeDoussal:2018dls,cafasso2019fredholm}.

The probability distribution \eqref{higher TW } can be used to compute the contribution of the edge fluctuation to the partition function of the generalized GWW model by using the Eqs. \eqref{pf prob} and \eqref{two-fold}. In fact, there are three regions in the vicinity of the edge, namely the finitely ($\cO(1)$) close regions in the left and right sides of the edge, and the crossing region which is infinitesimally close to the edge. 
At the edge, the dynamics is determined by the $p$-Airy process and the fluctuations in the right/left ($\pm$) sides of the edge contribute to the free energy \eqref{tot F}, up to a scaling factor $N^{-2}$, as 
\bea
\cF_f\sim\lim_{s\to \pm\infty} \log\det \left(1- K_{p\text{-Airy}}\right)_{(s,\infty)}.
\eea
Therefore, the right/left ($\pm$) edge free energy are obtained from the left and right tails of the higher TW distribution,
\bea
\cF_f\sim \lim_{s\to +\infty}\log F_p(s), \qquad \cF_f\sim \lim_{s\to -\infty}\log F_p(s).
\eea

In the special case of the Plancherel partition, using the scaled parameters $(x,\mathfrak{q}) \to (x/\epsilon,\mathfrak{q}/\epsilon)$, the difference equation~\eqref{eq:ODE_B} becomes
\begin{align}
 \qty[ \epsilon^2 \dv[2]{}{x} - \qty( \frac{x}{\mathfrak{q}} - 2 ) + \cO(\epsilon^4) ] J_{x/\epsilon}(2\mathfrak{q}/\epsilon) = 0
 \, ,
\end{align}
and in the scaling limit, the wave function becomes the Airy function,
\begin{align}
    J_{x/\epsilon}(2\mathfrak{q}/\epsilon)
    \ \xrightarrow{\epsilon \to 0} \
    \operatorname{Ai}(\xi)
    \, ,
\end{align}
where $\xi=\epsilon^{-2/3} \mathfrak{q}^{-1/3}(x-2 \mathfrak{q})$.
Similarly, the scaling limit of the discrete Bessel kernel is given by the Airy kernel,
\begin{align}
    K\qty(\frac{2\mathfrak{q}}{\epsilon} + \qty(\frac{\mathfrak{q}}{\epsilon})^{1/3} x, \frac{2\mathfrak{q}}{\epsilon} + \qty(\frac{\mathfrak{q}}{\epsilon})^{1/3} y)
     \xrightarrow{\epsilon \to 0} \
     \qty(\frac{\mathfrak{q}}{\epsilon})^{1/3} K_\text{Airy}(x,y)
    \, .
\end{align}

As a special case of the generic $p$, in the case of $p=2$, the dynamics of the model is governed by the Airy process and TW distribution, and the contribution of the fluctuation to the free energy is given, up to a scaling factor $N^{-2}$, by 
\bea
\cF_f\sim\lim_{s\to \pm\infty}\log\det \left(1- K_{\text{Airy}}\right)_{(s,\infty)}=\lim_{s\to \pm\infty}\log F_
2(p).
\eea

\section{Critical dynamics}\label{sec:crit_dyn}
In this Section, we study the free energy of the generic unitary matrix model from the viewpoint of the fluctuation of the model around the edge and explore the universal phase structure of the model in the vicinity of a critical point. As we will see in the following Sections this phase transition is associated with the opening/closing of a gap in the distribution function of the eigenvalues on the circle. From the mathematical point of view, the case $p=2$ explains the critical dynamics of the matrix models.
\subsection{General unitary matrix model and TW distribution}
Precisely speaking, in the particular case $p=2$ of the Eq.\eqref{higher TW }, and by assuming the scaling relation between $\epsilon$ and $N$, discussed in Section \ref{bulk fluc}, the finite and large $N$ results for the free energy of the critical model is encoded of the following result,
\begin{align}
    \lim_{\epsilon \to 0} \mathbb{P}\qty[ \frac{\lambda_1 - c_1 / \epsilon}{(c_2 / \epsilon)^{1/3}} < s] 
    =\det( 1 - K_{\text{Airy}})_{(s,\infty)}
    = F_2(s)
    \, ,
    \label{TW2}
\end{align}
where $c_1$ and $c_2$ are some model-dependent parameters which in the case of generalized GWW model we have $c_1= \beta(t_1, t_2, ...)$ and $c_2= \alpha_2(t_1, t_2, ...)$ in Eq.\eqref{eq:alpha_beta}, and  $F_2$ is the TW distribution and
\bea
K_{\text{Airy}}(x,y)=\frac{\text{Ai}(x) \text{Ai}'(y) - \text{Ai}(y) \text{Ai}'(x) }{x-y}.
\eea
Using the definition \eqref{F def}, and by fixing the 't Hooft parameter $\gamma := N \epsilon$ in the large $N$ limit of the $\mathrm{U}(N)$ matrix model, the universal result for the free energy can be obtained from \eqref{pf prob}, \eqref{tot F}, and \eqref{TW2}, up to a scaling factor $N^{-2}$, as
\begin{align}
    \mathcal{F}_f\sim\lim_{s \to \pm\infty} \log F_2(s)
    \qquad \text{where} \qquad
    s= \frac{\gamma-c_1}{c_2^{1/3}}N^{\frac{2}{3}},
\end{align}
and the free energy in the left side of the edge is given by
\bea
\cF= \cF_c+ \lim_{N\to \infty}N^{-2}\log F_2(s), \qquad \text{for} \qquad \gamma < c_1,\, s\to-\infty, 
\eea
and right sides of the edge is given by
\bea
\cF= \cF_c+ \lim_{N\to \infty}N^{-2}\log F_2(s), \qquad \text{for} \qquad \gamma> c_1, \, s\to+\infty.
\eea

In the rest of this section, we explain the perturbative and non-perturbative aspects of the general unitary matrix model in the weak and strong coupling phases, using the TW distribution. We obtain some results about the finite $N$ and genus expansion of the free energy. In the large $N$ limit, we compute the free energy and extract the phase structure of the models. Let us start with the asymptotic analysis of the TW distribution.

\subsubsection*{Tracy--Widom distribution}
The Tracy--Widom distribution is given by
\bea
F_2(s)= \exp{-\int_s^\infty (x-s)\ q^2(x)\ \dd x},
\label{TW}
\eea
where $q(x)$ is the solution of the Painlev\'{e} II equation,
\bea
q_{xx}(x) = 2\,q^3(x) + x \, q(x),
\label{painleve II}
\eea
and $q_{xx}$ denotes the second derivative of $q$ with respect to $x$.
The asymptotic behavior of the solution is obtained in \cite{deift1995asymptotics,hastings1980boundary}, and as $x\to -\infty$, 
\bea
q(x) = \sqrt{-\frac{x}{2}}\left(1+ \frac{1}{8x^3}-\frac{73}{128 x^6}+\frac{10219}{1024 x^9}+ \cO(|x|^{-12})\right),
\label{q asymptotic}
\eea
and
\bea
q\left(x\right)\to \text{Ai}(x), \qquad \text{as} \qquad x\to\infty.
\eea
Using the Hastings--McLeod results \cite{hastings1980boundary} we have
\bea
F_2(s) =\exp{-\int_s^\infty R(x)\ \dd x},\qquad \text{where} \qquad R(x) =\int_x^\infty q^2(s)\ \dd s.
\eea
Moreover, they obtained
\bea
R(x)= q_x(x)^2 - x\, q(x)^2 - q(x)^4.
\eea
Unlike the Gaussian distribution, TW distribution is an asymmetric distribution and has different left and right tails. 
Using the Hastings--McLeod results, asymptotic analysis of the TW distribution is performed in \cite{baik2008asymptotics}, and the following result is obtained,
\begin{equation}
 \label{asym TW}
 F_2(s) = 
  \begin{cases}
  \displaystyle
   1-\frac{e^{-\frac{4}{3}s^{\frac{3}{2}}}}{32\pi s^{\frac{3}{2}}}\qty(1-\frac{35}{24s^{\frac{3}{2}}})+\mathcal{O}(s^{-3})                       & \quad s \rightarrow \infty \\[1em]
   \displaystyle
   2^{\frac{1}{24}}\, e^{\zeta'(-1)}\frac{e^{-\frac{1}{12}|s|^{3}}}{|s|^{\frac{1}{8}}} \qty(1-\frac{3}{2^{6}\,|s|^{3}}+\mathcal{O}(s^{-6}) )     & \quad s\rightarrow -\infty \\
  \end{cases},
\end{equation}
where $\zeta(x)$ is the Riemann zeta function. The asymmetry of the tails of the TW distribution is apparent in the leading order asymptotic,
\begin{equation}
\label{ATW}
F_2(s) =
  \begin{cases}
   1-\mathcal{O}(e^{-s^{3/2}}) & \quad s \rightarrow \infty \\
   \\
   \mathcal{O}(e^{-|s|^{3}}) & \quad s \rightarrow -\infty \\
  \end{cases}.
\end{equation}

\subsubsection{Finite $N$ results and $1/N$ expansion}
\label{1/N expansion}
Let us define the double-scaling parameter,
\bea
s=\alpha_2^{-\frac{1}{3}} (\beta_c-\beta)\ N^{\frac{2}{3}}.
\eea 
By using the double scaling parameter
for general unitary matrix model, the leading and sub-leading contributions to the free energy, can be obtained using the asymptotic expansion of the TW distribution \eqref{asym TW}, 
\begin{align}
    \cF = \begin{cases}
    \displaystyle
        \cF_c+ \frac{1}{N^2}\log\left(1- (32\pi)^{-1} s^{-\frac{3}{2}}\exp{-\frac{4}{3}s^{\frac{3}{2}}}\right) &  \beta<\beta_c, \ |\beta-\beta_
        c|=\cO(1)\\[1em]  \displaystyle
        \cF_c+\frac{1}{N^2}\log\left(2^{\frac{1}{24}}e^{\zeta'(-1)}e^{-\frac{1}{12}|s|^{3}}|s|^{-\frac{1}{8}} \left(1-\frac{3}{2^{6}\,|s|^{3}}\right)\right)& \beta>\beta_c, \ |\beta-\beta_
        c|=\cO(1) 
        \end{cases}.
\end{align}
We can add the crossing region, between the two tails, to the above result and after expanding the logarithm, we obtain
\begin{align}
    \cF = \begin{cases}
    \displaystyle
        \cF_c- \frac{1}{N^2}\sum_{n=1}^\infty\frac{1 }{n}\, (32\pi)^{-n}s^{-\frac{3}{2}n}\exp({-\frac{4}{3}n s^{\frac{3}{2}}}) & \beta<\beta_c, \ |\beta-\beta_
        c|= \cO(1)\\[1em] \displaystyle
        \cF_c +\frac{1}{N^2}\int_{s}^{\infty} (x-s)\ q^2(x)\ \dd x  &  \beta<\beta_c, \ |\beta-\beta_
        c|= \cO(N^{-2/3})\\[1em] \displaystyle
        \cF_c+\frac{1}{N^2}\left(c-\frac{|s|^3}{12} -\frac{1}{8} \log |s|-\sum_{n=1}^\infty \frac{1}{n}\left(\frac{3}{2^{6}}\right)^n |s|^{-3n}\right)   &  \beta>\beta_c, \ |\beta-\beta_
        c|= \cO(1) 
        \end{cases},
\end{align}
where $c=\frac{\log 2}{24}+ \zeta'(-1)$.
We can recast the expansions as
\begin{align}
    \cF = \begin{cases}
    \displaystyle
        \cF_c+ \sum_{n=1}^\infty N^{-n-2}\ G_n^{\ (1)}\ e^{-2 N f_n(\beta)} & \beta<\beta_c, \ |\beta-\beta_
        c|= \cO(1)\\[1em] \displaystyle
        \cF_c +\sum_{n=0}^\infty N^{-\frac{2}{3}n-2}\ G_n^{\ (2)}  &  \beta<\beta_c, \ |\beta-\beta_
        c|=\cO(N^{-2/3})\\[1em] \displaystyle
        \cF_c+ \cF_0 + N^{-2}\cG  -\sum_{n=1}^\infty N^{-2n-2}\ G_n^{\ (3)} &  \beta>\beta_c, \ |\beta-\beta_
        c|= \cO(1)
        \end{cases},
        \label{critical 1/N}
\end{align}
where in the right tail, the first line, we have
\bea
G_n^{\ (1)}=-\frac{1}{n} (32\pi)^{-n} \alpha_2^{\frac{n}{2}}\, (\beta_c-\beta)^{-\frac{3n}{2}},\qquad f_n(\beta)=\frac{2}{3}n\, \alpha_2^{-\frac{1}{2}}\ (\beta_c-\beta)^{\frac{3}{2}},
\label{G1 and f}
\eea
and in the left tail, third line,
\bea
\cF_0 =-\frac{\alpha_2^{-1}}{12} |\beta_c-\beta|^3, \quad \cG = \frac{\log 2}{24}+ \zeta'(-1) +\frac{1}{24}\log \alpha_2-\frac{1}{12}\log N -\frac{1}{8} \log |\beta_c-\beta|,\nonumber\\
\eea
and
\bea 
 G_n^{\ (3)}=\frac{1}{n}\left(\frac{3}{2^{6}}\right)^n\ \alpha_2^{n}\ |\beta_c-\beta|^{-3n}.
 \label{G(3)}
\eea
The above result reproduces the known results about the scaling behavior of the free energy of GWW model \cite{liu2004fine} and moreover generalize it to the generalized GWW model \cite{alvarez2006black}. We will elaborate more on this in Sections \ref{sec: Plancherel partition and GWW} and \ref{sec: Schur measure and generalized GWW}.
\subsubsection*{Right tail of TW distribution and instantons}
The right tail of the TW distribution is obtained by taking the limit $s\to \infty$, thus we can expand the exponential in Eq.\eqref{TW}, as the integrand becomes small in this limit, 
\bea
F_2(s)\approx 1-\int_s^\infty (x-s)\ q^2(x)\ \dd x,
\label{F2 approx}
\eea
then, using the asymptotic result
\bea
q\left(s\right)\to \text{Ai}(s),\quad \text{if} \quad s\to\infty,
\eea
and the Airy function asymptotics,
\bea
\text{Ai}(s)= \frac{1}{2\sqrt{\pi}s^{1/4}}\exp{-\frac{2}{3}s^{\frac{3}{2}}}\left(1-\frac{5/27}{\frac{2}{3} s^{\frac{3}{2}} }+ \cO\left(\frac{1}{|s|^3}\right)\right),
\eea
by keeping the first term in the above expansion of the Airy function, and using it in Eq.\eqref{F2 approx}, we obtain
\bea
F_2(s)\approx 1-\frac{1}{4\pi}\int_s^\infty (x-s)\ x^{-\frac{1}{2}}\exp{-\frac{4}{3}x^{\frac{3}{2}}}\ \dd x.
\eea
The above integral can be evaluated as
\bea
F_2(s)\approx 1-\frac{1}{8\pi} \exp{-\frac{4}{3}s^{\frac{3}{2}}}+\frac{1}{6\pi}s^{\frac{3}{2}} E_{\frac{2}{3}}\left(\frac{4}{3}s^{\frac{3}{2}}\right),
\eea
where $E_n(x)$ is the generalized exponential integral defined by
\bea
E_n(x)= \int_1^\infty \frac{e^{-xt}}{t^n} \ \dd t.
\eea
Using the asymptotic expansion of the generalized exponential function at $x \to \infty$, 
\bea
E_n(x)= \frac{e^{-x}}{x}\left(1-\frac{n}{x}+ \frac{n(n+1)}{x^2}- ...\right),
\label{expan gen expon integ}
\eea
we obtain
\bea
F_2(s) \approx 1- (16\pi)^{-1} s^{-\frac{3}{2}}\exp{-\frac{4}{3}s^{\frac{3}{2}}}\left(1-\frac{5}{4 s^{\frac{3}{2}}}+\cO(s^{-3})\right).
\eea
The above right tail expansion, $s\to\infty$ of the TW distribution, matches, up to some numerical coefficients, with the similar expansion in the literature, for example in \cite{baik2008asymptotics,borot2012right},
\bea
F_2(s) = 1- (32\pi)^{-1} s^{-\frac{3}{2}}\exp{-\frac{4}{3}s^{\frac{3}{2}}}\left(1-\frac{35}{24 s^{\frac{3}{2}}}+\cO(s^{-3})\right).
\label{eq: right tail TW2}
\eea
\subsubsection*{Noperturbative instanton sector}
In the region $\beta>\beta_c, \ (\beta-\beta_c)= \cO(1)$, using the parameterization $s= (\xi N)^{\frac{2}{3}}$, we observe that there is no perturbative corrections to the continuum free energy and the non-perturbative corrections due to the instantons are of order $\cO(e^{-N})$, and we can expand the leading terms of Eq.\eqref{eq: right tail TW2} to obtain the instanton contribution to the free energy,
\bea
\cF_{\text{inst}}=- \sum_{n=1}^\infty\frac{(32\pi\xi)^{-n} }{n} N^{-n-2}\exp(-\frac{4}{3}n\, \xi N),
\label{inst free energy}
\eea
where $n$ is the number of the instantons. Thus, we can write $\displaystyle \cF_{\text{inst}}=\sum_{n=1}^\infty \cF_{n\text{-inst}}$, and in particular, the one-instanton sector, which is the dominant contribution, is
\bea
\cF_{1\text{-inst}}= -\frac{1}{32\pi\xi} N^{-3}\exp(-\frac{4}{3}\ \xi N).
\eea
\subsubsection*{Left tail of TW distribution and genus expansion of free energy}
In this part, we use the left tail expansion of the TW distribution to obtain the genus expansion of the free energy.
The genus expansion of the free energy is defined by
\bea
\cF=\sum_{g=0}^\infty N^{2-2g} \cF_g.
\label{genus zero F}
\eea
Notice that in the genus expansion of the free energy, the definition of the free energy is not normalized by the $1/N^2$ factor and for example the genus zero free energy is of order $N^2$, and this should be considered when we compare the genus expansion with the result in Section \ref{1/N expansion}.
In the genus expansion of the free energy, for more convenience, we use slightly different parameterization $s= \lambda N^{\frac{2}{3}}$, in which the two parameters are related by $\lambda= \xi^{\frac{2}{3}}$. In this parameterization, we can arrange the expansion of the free energy in powers of $N$ and find the subleading terms in $1/N$ expansion. Using this parameterization we find the total genus zero part, which is defined as the sum of the continuum free energy and genus zero free energy in the expansion \eqref{genus zero F}, as 
\bea
\tilde\cF_0=\cF_c+\cF_0&=& \cF_c +\begin{cases}
    \displaystyle
        0 & \text{for } \beta<\beta_c\\[1em] \displaystyle
        -\frac{1}{12} |\lambda|^3 & \text{for } \beta>\beta_c 
        \end{cases}.
\eea
Furthermore, as we observed, there is no perturbative corrections $\cO(\frac{1}{N})$ in the region $\beta>\beta_c$, and the perturbative corrections in the region $\beta<\beta_c$, can be computed by using the left tail asymptotics of the $\text{TW}^{(2)}$ distribution.
In fact, the free energy at genus one, and for higher genus ($g\geq2$), are given by
\bea
\cF_1&=&\frac{\log 2}{24}+ \zeta'(-1) -\frac{1}{8}\log|\lambda|- \frac{1}{12}\log N,\qquad  \cF_g= -\frac{1}{g-1}\left(\frac{3}{2^{6}}\right)^{g-1} \lambda^{3-3g}.
\eea
Let us define the perturbative corrections to free energy by
\bea
\cF_{\text{pert}}=\cF-\cF_0=  \sum_{g=1}^\infty N^{2-2g} \cF_g,
\label{pert free energy}
\eea
and then by putting the perturbative and non-perturbative leading and subleading terms together in the two phases, we obtain
\bea
        \cF=\begin{cases}
    \displaystyle
        \cF_c+\cF_{\text{inst}}, & \text{for } \beta<\beta_c\\[1em] \displaystyle
       \cF_c-\frac{1}{12} |\lambda|^3+\cF_{\text{pert}}, & \text{for } \beta>\beta_c
        \end{cases}.
        \label{free energy gen}
\eea
Notice that, the above result for the free energy is universal, besides the model-dependent parts, namely the continuum free energy $\cF_c$, and parameters $\xi$ and $\lambda$.

\subsubsection{Large $N$ asymptotics: Universal phase structure}
As we observed so far, the different asymptotic behavior of the TW distribution in its left and right tail, lead to different expansion of the free energy in the strong and weak coupling phases. In fact, there is a universal strong-to-weak coupling transition, i.e. the deconfinement phase transition, in the unitary matrix models.
The large gap asymptotic of the Fredholm determinant, which is the asymptotic analysis in the left tail (i.e. the phase $\beta > \beta_c$) of the TW distribution,
 is obtained via the Riemann--Hilbert analysis \cite{baik2008asymptotics}, as $s\to -\infty$,
\bea
F_2(s)= 2^{1/24} e^{\zeta'(-1)}\, |s|^{-\frac{1}{8}}\exp{-\frac{|s|^3}{12}}\left(1+\frac{3}{2^6 |s|^3}+ \cO(|s|^{-6})\right).
\label{F2 exp corr}
\eea
Thus, we obtain the universal finite contribution from the edge fluctuation to the free energy in strongly-coupled phase,
\bea
\cF_f\sim\lim_{s\to-\infty} \log F_2(s)=-\frac{|s|^3}{12} -\frac{1}{8} \log |s| + \log C_2 + o(1),  
\eea
where $s=\alpha_2^{-\frac{1}{3}} (\beta_c-\beta)\ N^{\frac{2}{3}}$, and the parameters $\alpha$ and $\beta$ depend on the model, and $C_2=2^{1/24} e^{\zeta'(-1)}$.

Putting together the left and right tails expansion of the TW distribution, the leading order free energy and the order of the subleading perturbative and non-perturbative corrections, can be summarized as
\bea
        \cF=\begin{cases}
    \displaystyle
        \cF_c+\cO(e^{-cN}) & \text{for } \beta<\beta_c\\[1em] \displaystyle
       \cF_c-\frac{1}{12} |\lambda|^3+\cO(N^{-2}) & \text{for } \beta>\beta_c
        \end{cases}.
\eea

It is easy to observe that the third derivative of the free energy is discontinuous at $\beta(t_i) =\gamma\equiv\beta_c$, which is leading to a universal third order phase transition in the model.  This is a universal generalization of the GWW phase transition at $\beta=\beta_c$ between the two phases: weak coupling $\beta < \beta_c$ ($s\to \infty$) and strong coupling $\beta > \beta_c$ ($s\to -\infty$). 
This is the phase transition associated with the one-gap opening/closing in the unitary matrix model. 

\subsubsection*{Cross-over region between the tails}
Having discussed the left and right tails of the TW distribution, in this Section, we take a closer look on the crossing regions between the tails. At finite but large $N$, the free energy of the matrix model in the intermediate region is obtained from the TW distribution,
\bea
\cF= \cF_c - \int_{s}^{\infty} (x-s)\ q^2(x)\ \dd x,
\eea
where $s$ is finite, $s\sim \alpha_2^{-1/3}=\cO(1)$, as $\beta-\beta_c= \cO(N^{-2/3})$, and $q(x)$ is the solution of the Painlev\'e II equation \eqref{painleve II}. 
Notice that, from Eq.\eqref{TW}, one can observe that $q$ satisfies
\bea
q^2(s)= -\dv[2]{s} \log F_2(s),
\eea
and one can interpret $q^2$ as the specific heat, as $F_2$ has the interpretation of the partition function in our context. 

In order to approximate the $F_2(s)$ in the crossing region for finite $s$, one needs to study the interpolation of the function $q(x)$ between the left and right tails; in the limit $x\to -\infty$, the asymptotic behavior is given by the Eq.\eqref{q asymptotic}
and in the limit $x\to\infty$, it is $q\left(x\right)\to \text{Ai}(x)$.
In the crossing region, it is numerically obtained that the TW distribution can be approximated by the Gamma distribution \cite{chiani2014distribution},
\bea
F_2(s)\approx \Gamma(k)^{-1} \gamma(k,\frac{s+\alpha}{\theta}),
\eea
where $\gamma(k,x)$ is the lower incomplete gamma function defined as
\bea
\gamma(t,x)=\int_{0}^{x} s^{t-1} e^{-s}\, \dd s, 
\eea
and the parameters $k$, $\alpha$ and $\theta$ are numerically adjusted to fit the TW distribution.

Having discussed the applications of the asymptotic analysis of the TW distribution in the general setting, in the rest of this Section, we apply our result in two concrete example of the GWW model and its generalization.
\subsection{GWW model and TW distribution}\label{GWW section}
\label{sec: Plancherel partition and GWW}
The perturbative and non-perturbative results and a universal phase structure in the GWW matrix model can be extracted from the free energy given by
\begin{align}
    \mathcal{F}=\cF_c+\lim_{N \to \infty} N^{-2} \log F_2(s),
    \qquad \text{where} \qquad
    s= \lambda N^{2/3},
    \label{F GWW}
\end{align}
and $\lambda= \alpha_2^{-1/3}(\beta_c-\beta) $, with the coefficients $\alpha_2$ and $\beta$, given by
\begin{align}
    \alpha_2 =  t_1 
    \, , \qquad
    \beta = 2 t_1
    \, , \qquad
    \beta_c=\gamma.
    \label{}
\end{align}
Assuming $\gamma=1$, the critical point $\beta_c=1$ implies that the critical coupling is $t_1^*=1/2$. Then, we easily obtain
\begin{align}
    \lambda =  \frac{1-2t_1}{t_1^{1/3}}
    \, , \qquad
    \xi = \frac{(1-2t_1)^{3/2}}{t_1^{1/2}}   ,
    \label{}
\end{align}
and we can compute the free energy \eqref{free energy gen}, 
\bea
\cF=\begin{cases}
    \displaystyle
        t_1^2+\cF_{\text{inst}} & \text{for } t_1<1/2\\[1em] \displaystyle
       t_1^2-\frac{|1-2t_1|^3}{12 t_1}+\cF_{\text{pert}} & \text{for } t_1>1/2
        \end{cases},  
        \label{GWW F}
\eea
where $\cF_{\text{inst}}$ in Eq.\eqref{inst free energy} can be computed by sum over all $n$-instantons contributions, given by 
\bea
\cF_{n\text{-inst}}=\frac{1}{(32\pi)^n n} \frac{t_1^{n/2}}{(1-2t_1)^{3n/2}}  N^{-n-2}\exp(-\frac{4}{3} n\ \frac{(1-2t_1)^{3/2}}{t_1^{1/2}} N),
\eea
and $\cF_{\text{pert}}$ is given by
\bea
\cF_1=\frac{\log 2}{24}+ \zeta'(-1) -\frac{1}{8}
    \log \left|\frac{1-2t_1}{t_1^{1/3}}\right|- \frac{1}{12}\log N.
     \label{F genus one t1}
\eea
and 
\bea
\cF_g= -\frac{1}{g-1}\left(\frac{3}{2^{6}}\right)^{g-1}  \left|\frac{1-2t_1}{t_1^{1/3}}\right|^{3-3g},\quad \text{for } g\geq 2.
\label{F pert t1}
\eea

To compare with results in the literature for example \cite{liu2004fine,okuyama2017wilson}, notice that the coupling of our GWW matrix model $t_1$ is related to one in the literature $t$,  by $t=2t_1$. 
The leading term of the free energy of the GWW model obtained from the Coulomb gas method is
\bea
\cF=\begin{cases}
    \displaystyle
        \frac{t^2}{4} & \text{for } t<1\\[1em] \displaystyle
       t -\frac{3}{4} - \frac{\log t}{2} & \text{for } t>1
        \end{cases}.
        \label{GWW F CG}
\eea
Above free energy implies there is a third order phase transition at $t=1$ in GWW model. This model-dependent third-order phase transition is an example of the universal third order phase transition in our discussion implied by the universality of the TW distribution. 
In order to compare the above model dependent result with the leading terms (genus zero) of the universal result obtained in Eq.\eqref{GWW F}, we should consider the vicinity of the critical point $t^*=1$. The free energy obtained by both methods in the phase $t>1$ matches exactly. In the phase $t<1$ and in the vicinity of the critical point $t\nearrow 1$, one can observe that the expansion of the free energy in terms of $\epsilon = t-1$ (or $t=e^\epsilon$) in Eqs.\eqref{GWW F} and \eqref{GWW F CG} matches up to order $\cO(\epsilon^3)$.

Let us consider the $1/N$ expansion in GWW model \cite{liu2004fine},
\begin{align}
    \cF = \begin{cases}
    \displaystyle
        \frac{t^2}{4}+ \frac{1}{2\pi}e^{-2 N f(t)}\sum_{n=1}^\infty N^{-n-2}\ F_n^{\ (1)}\  & \text{for } t<1,\ 1-t= \cO(1) \\[1em] \displaystyle
        \frac{t^2}{4} +\sum_{n=0}^\infty N^{-\frac{2}{3}n-2}\ F_n^{\ (2)}  & \text{for } t<1, \ 1-t= \cO(N^{-2/3})\\[1em] \displaystyle
         t -\frac{3}{4} - \frac{\log t}{2}  +\sum_{n=0}^\infty N^{-2n-2}\ F_n^{\ (3)} & \text{for } t>1,\ t-1= \cO(1)
        \end{cases},
\end{align}
where 
\bea
f(t)= \log\left(\frac{1}{t}+ \sqrt{\frac{1}{t^2}-1}\right) - t\sqrt{\frac{1}{t^2}-1},\quad F_n^{\ (1)}\sim \frac{1}{(1-t)^{\frac{3n}{2}}}, \quad F_n^{\ (3)}\sim \frac{1}{(t-1)^{3n}},\nonumber\\
\eea
and $F_0^{\ (2)}$ satisfies Painlev\'{e} II equation \cite{liu2004fine}. We observe an agreement between $G_n^{(1,2,3)}$ in Eq.\eqref{critical 1/N} and $F_n^{(1,2,3)}$ in above result.
In instanton sector, different approximation for Airy function is applied in our study whereas in the literature the Airy function is approximated by the Bessel function and its Debye approximation. However, one can observe that the $f_1(\beta)$ in Eq.\eqref{G1 and f} matches with $f(t)$ around $t=1$ up to the first term in the expansion which is of order $\cO(\epsilon^{3/2})$. We hope to comeback to this problem, in our future studies.

The perturbative genus expansion in the GWW model \cite{okuyama2017wilson} is
\bea
\cF_g(t)= \frac{B_{2g}}{2g(2g-2)}+ \frac{1}{(t-1)^{3g-3}} \sum_{n=0}^{g-2} c_n^{(g)} t^n,
\label{genus exp GWW}
\eea
where $B_{2g}$ denotes Bernoulli number. 
In particular, at genus one, we have
\bea
\cF_1(t)=\zeta'(-1) -\frac{1}{8}
    \log (1-1/t)- \frac{1}{12}\log N.
    \label{genus 1 GWW}
\eea
In general, for any $g\geq 1$, we expect that the above results match with Eqs.\eqref{F genus one t1} and \eqref{F pert t1} around the critical point $t=1$. At genus one, all terms in Eq.\eqref{genus 1 GWW} match with Eq.\eqref{F genus one t1} up to the coefficient of term $\log t$. 
The first term in Eq.\eqref{genus exp GWW} is normalization for the Haar measure and it is obtained by computing the volume of the $\mathrm{U}(N)$ gauge group. However, for the second term in Eq.\eqref{genus exp GWW}, by comparing with Eq.\eqref{F pert t1}, although they are different in order, but we expect the following relation around $t=1$, 
\bea
\sum_{n=0}^{g-2} c_n^{(g)} t^n \sim -\frac{1}{g-1}\left(\frac{3}{2^{7}}\right)^{g-1} t^{g-1}.
\eea
More precise comparison and further implications of the above equation remain for future studies.

\subsection{Generalized GWW model and TW distribution}
\label{sec: Schur measure and generalized GWW}
In this Section we explain the free energy of the one-cut solution of the generalized GWW model.
Similar to GWW model, all the results in the matrix model with an arbitrary potential in generalized GWW can be obtained from the free energy \eqref{F GWW} with explicit model-dependent parameters given by
\begin{align}
    \alpha_2(t_1, t_2, ...) = \sum_{n=1}^\infty n^{3} \, t_n 
    \, , \qquad
    \beta(t_1, t_2, ...)= 2\sum_{n=1}^\infty n \, t_n 
    \, , \qquad
    \beta_c(t_1, t_2, ...)=\gamma
    \, .
    \label{gGWW parameters}
\end{align}
Then, by using explicit form of $\lambda$ and $\xi = \lambda^{3/2}$ 
in Eqs.\eqref{inst free energy}, and \eqref{pert free energy}, one can explicitly compute the free energy
of the generalized GWW model,
\bea
    \cF = \sum_{n} n\, t_n^2+ \begin{cases}
    \displaystyle
        \cF_{\text{inst}} & \text{for } \beta<\beta_c\\[1em] \displaystyle
       -\frac{1}{12} |\lambda|^3+\cF_{\text{pert}} & \text{for } \beta>\beta_c
        \end{cases}.
        \label{F gGWW}
\eea
The above free energy of the generalized GWW model implies a universal third order phase transition in the generalized GWW model.
In the following, we compare our result with $1/N$ expansion of the free energy in the literature.
The critical behaviour of the generalized GWW model is discussed in~\cite{alvarez2006black},
\begin{align}
    \cF = \begin{cases}
    \displaystyle
        \sum_n n \ t_n^2+ \frac{1}{2\pi}e^{-2 N f(t_n)}\sum_{n=1}^\infty N^{-n-2}\ F_n^{\ (1)}\  & \text{for } t<1,\ 1-t= \cO(1) \\[1em] \displaystyle
        \sum_n n \ t_n^2 +\sum_{n=0}^\infty N^{-\frac{2}{3}n-2}\ F_n^{\ (2)}  & \text{for } t<1, \ 1-t= \cO(N^{-2/3})\\[1em] \displaystyle
         H(t_n)  +\sum_{n=0}^\infty N^{-2n-2}\ F_n^{\ (3)} & \text{for } t>1,\ t-1= \cO(1)
        \end{cases},
\end{align}
where $F_n^{\ (1,2,3)}$ and $H(t_n)$ can be computed using the methods of orthogonal polynomials~\cite{goldschmidt19801}.
In the vicinity of the critical point, assuming $\gamma=1$,
\bea
2\sum_n n\, t_n^*=1,
\eea
using the results in Eqs.\eqref{critical 1/N}-\eqref{G(3)} with the parameters $\alpha_2$ and $\beta$ given by Eq.\eqref{gGWW parameters}, in principle we can obtain explicit results for $H(t_n)$, $f(t_n)$, $F_n^{(1,3)}$ and $F_0^{(2)}$. Similar to GWW case, they can compared with $G_n^{1,2,3}$ in Eqs. \eqref{G1 and f} and \eqref{G(3)} with insertion of parameters from Eq.\eqref{gGWW parameters}. In particular, the genus zero free energy in the strong coupling regime is expected to be obtained from Eq.\eqref{F gGWW}, as
\bea
H(t_n) \approx  \sum_n n\, t_n^2  -\frac{1}{12} \alpha_2^{-1}|1-\beta|^3.
\eea
In summary, the new results include the exact and explicit results for the $1/N$ expansion, genus expansion at finite $N$ and phase structure of the model at large $N$.

As a simple example, we can consider the single-term generalized GWW model, $t_n=t_m\delta_{n,m}$, with the following potential
\bea
V(U)= t_m \tr(U^{m}+ U^{-m}).
\eea
For simplicity we write $t_m\equiv t$ and then in this case we obtain the double scaling parameter
\bea
\lambda= m^{-1} t^{-\frac{1}{3}} (1-2m t),
\eea
and the free energy,
\bea
    \cF = m\, t^2+ \begin{cases}
    \displaystyle
        \cF_{\text{inst}} & \text{for } t<\frac{1}{2m}\\[1em] \displaystyle
       -\frac{1}{12} m^{-3} t^{-1}|1-2mt|^3 +\cF_{\text{pert}} & \text{for } t>\frac{1}{2m}
        \end{cases}.
\eea
This implies that there is a third-order phase transition at critical coupling $t^*= \frac{1}{2m}$. Similarly, $\cF_{\text{inst}}$ and $\cF_{\text{pert}}$ can be computed explicitly.
\section{Multi-critical dynamics}
\label{multi section}



In the previous Section, we have studied the critical dynamics associated with the one gap opening/closing in the GWW model and its generalization. In this Section, we study the multi-critical extension of our previous results for the generalized GWW model, which are related to the multi-gap opening/closing in the unitary matrix model with a generic high degree polynomial potential. The multi-critical dynamics is originated in the asymptotic behavior of the higher TW distribution $\text{TW}^{(p)}$. The asymptotic behavior of the TW distribution ($p=2$) is studied in \cite{baik2008asymptotics}, and recent asymptotic results for the Pearcey processes
($p=3$) are obtained in \cite{dai2020asymptotics}.
See also earlier results~\cite{Brezin:1998zz,Brezin:1998PREb}. The asymptotic analysis for the generic case of higher $\text{TW}^{(p)}$ distribution is conjectured in \cite{LeDoussal:2018dls} and proved in \cite{cafasso2019fredholm}. See also ~\cite{Periwal:1990qb,Claeys:2009CPAM}. In this Section we review recent results in the large gap asymptotics of the higher TW distributions. Moreover, we obtain the asymptotics results for the right tail expansion. Then, we apply these results in the multi-critical dynamics of the unitary matrix model. The analysis of this Section is the direct generalization of the one for $p=2$ in previous Section.

\subsection{Multi-critical dynamics and higher $\text{TW}^{(p)}$ distribution}
The main result of this work is the critical/multi-critical phase structure of the generalized GWW model. In this study, we consider $p\in 2\mathbb{N}$, however we expect similar results for odd $p$ \cite{kimura2021}. We conjecture that all the perturbative and non-perturbative information at the finite $N$ and large $N$ and the (multi-)critical phase structure can be encoded in a compact form as
\begin{align}
    \lim_{\epsilon \to 0} \mathbb{P}\qty[ \frac{\lambda_1 - c_1^{(p)} / \epsilon}{(c_2^{(p)} / \epsilon)^{\frac{1}{p+1}}} < s] 
    = F_p(s)
    \, ,
    \label{TWp}
\end{align}
where $c_1^{(p)}$ and $c_2^{(p)}$ are some model dependent parameters, and $F_p(s)$ is 
$\text{TW}^{(p)}$ distribution.
In the generalized GWW model, we have the distribution \eqref{higher TW } with parameters are explicitly  given by
\begin{align}
    c_2^{(p)}=\alpha_p(t_1, t_2, ...) = 2\sum_{n=1}^\infty \frac{n^{p+1}}{p!} \, t_n 
    \, , \qquad
    c_1^{(p)}=\beta(t_1, t_2, ...) = 2\sum_{n=1}^\infty n \, t_n 
    \, .
    \label{alpha beta p}
\end{align}
The free energy in the large $N$ limit of the $\mathrm{U}(N)$ matrix model is defined in \eqref{F def}, and it can be obtained from \eqref{pf prob} and \eqref{TWp}, by fixing $\gamma := N \epsilon$, as
\begin{align}
    \mathcal{F}=\cF_c+\lim_{N \to \infty} N^{-2} \log F_p(s)
    \qquad \text{where} \qquad
    s= \frac{\gamma-c_1^{(p)}}{\left(c_2^{(p)}\right)^{\frac{1}{p+1}}}N^{\frac{p}{p+1}},
\end{align}
is called the multi-critical double-scaling parameter.
\subsubsection*{Multi-critical double-scaling parameter}
Let us 
define the critical point $\gamma=\beta_c$ and by using \eqref{alpha beta p}, write double scaling parameter in generalized GWW model as
\bea
s=\alpha_p^{-\frac{1}{1+p}} (\beta_c- \beta)\ N^{\frac{p}{p+1}}.
\label{p-scaling}
\eea
The multi-critical dynamics and phase structure is governed by the behaviour of free energy around the critical point $\beta=\beta_c$, and as we will see, the free energy in the vicinity of the critical point and the order of the multi-critical phase transition can be explicitly computed from the parameters $\alpha_p$ and $\beta$ in the asymptotic expansion of $F_p(s)$ around the critical point.
For real positive $t_n$, we have $\alpha_p>0,\ \beta>0$, and as $N\to \infty$, the double-scaling parameter $s$ is tending to plus/minus infinity depending on the sign of $(\beta_c-\beta)$ and thus there are two phases $\beta_c>\beta$ and $\beta_c<\beta$ in the matrix model. Then, the free energy has different expansions in different phases. Thus, to obtain the free energy one needs to study the asymptotics of $F_p(s)$ in the two limits of $s\to\pm\infty$.

\subsubsection{Multi-critical instantons}
In this Section, we study the right tail asymptotics of $\text{TW}^{(p)}$ and its applications for the  instantons in the multi-critical model. 

In the limit $s\to \infty$, one can obtain the asymptotic expansion of $F_p(s)$, in a straightforward calculations, by using the definition 
\bea
\log F_p(s)= -\int_{s}^{\infty} (x-s)\ q^2\left((-1)^{\frac{p}{2}+1}x\right) \dd x,
\label{TW-p}
\eea
and the following asymptotic result \cite{cafasso2019fredholm},
\bea
q\left((-1)^{\frac{p}{2}+1}s\right)\to \text{Ai}_{p}(s),\quad \text{if} \quad s\to\infty,
\eea
where $\text{Ai}_{p}(s)$ is the  $p$-Airy function defined in \eqref{eq:p-Airy_lim}.

The asymptotic behavior of the $p$-Airy function is obtained in \cite{bothner2021momenta}, and the following results, up to some unkown  numerical factor $a_p$ and $\tilde{a}_p$, holds
\bea
\text{Ai}_{p}(s)= a_p\, s^{-\frac{p-1}{2p}} \exp{-\frac{p}{p+1}s^{\frac{p+1}{p}}}\left(1-\tilde{a}_p\, s^{-\frac{p+1}{p}}+ \cO(s^{-\frac{2(p+1)}{p}})\right).
\label{higher Airy asymp}
\eea
The factor $\tilde{a}_p$ does not appear in the following approximation as we keep the leading term in the asymptotic, and we will also drop the factor $a_p$ in the following computation.
Similar to the discussion in the $p=2$ case, by using the leading term in the asymptotic of the higher Airy function \eqref{higher Airy asymp}, from Eq.\eqref{TW-p} in the limit $s\to \infty$, we obtain
\bea
F_p(s)\approx 1-\frac{1}{2} \exp{-\frac{2p}{p+1}s^{\frac{p+1}{p}}}+\frac{p}{p+1}s^{\frac{p+1}{p}} E_{\frac{p}{p+1}}\left(\frac{2p}{p+1}s^{\frac{p+1}{p}}\right).
\label{higher TW asymp 1}
\eea
Then, by expanding the generalized exponential integral Eq.\eqref{expan gen expon integ}, we obtain
\bea
F_p(s)\approx 1-\frac{1}{4} s^{-\frac{p+1}{p}}\exp{-\frac{2p}{p+1}s^{\frac{p+1}{p}}} \left(1-\frac{2p+1}{p}s^{-\frac{p+1}{p}} 
+\cO(s^{-\frac{2(p+1)}{p}})\right).
\label{higher TW asymp 2}
\eea
Finally, for generic $p$, we have the following right tail expansion of the logarithm of the higher $\text{TW}^{(p)}$,
\begin{align}
     \log F_p(s) \approx
    \log \left( 1 - 4^{-1} s^{-\frac{p+1}{p}} \exp({-\frac{2p}{p+1}s^{\frac{p+1}{p}}}) \right)=-\sum_{n=1}^\infty\frac{1 }{4^n n}s^{-\frac{n(p+1)}{p}}\exp({-\frac{2n p}{p+1}s^{\frac{p+1}{p}}}).
    \label{higher TW asymp 3}
\end{align}

The right tail of $\text{TW}^{(p)}$, explain the regime $\beta<\beta_c$, and $(\beta-\beta_c)= \cO(1)$ of the generalized GWW model. Using the parameterization $s= (\xi_p N)^{\frac{p}{p+1}}$, and the above asymptotic results, the free energy in this phase is obtained as
\bea
\cF=\cF_c+ \cF_{\text{inst}}^{(p)},
\eea
where 
\bea
\cF_{\text{inst}}^{(p)}= -\sum_{n=1}^\infty\frac{1 }{4^{n} n}\xi_p^{-n} N^{-n-2}\exp{-\frac{2n p}{p+1}\xi_p N}.
\label{multi inst}
\eea
In the generalized GWW model, we have $\cF_c=\sum_{n} n\, t_n^2$, and $\xi_p=\alpha_p^{-\frac{1}{p}} (\beta_c- \beta)^{\frac{p+1}{p}}$.
\subsubsection{Multi-critical genus expansions}
In this part, we study the left tail asymptotics of $\text{TW}^{(p)}$ and its application in the strong coupling phase of multi-critical model, such as the computation of the genus-expansion of the free energy.
The asymptotic expansion of the $\text{TW}^{(p)}$ distribution, in the limit $s\to -\infty$, is recently obtained in \cite{cafasso2019fredholm}, 
\bea
\log F_p(s)= c_p |s|^{\frac{2(p+1)}{p}} + c \log |s| + \log C_p + o(1),
\label{TWp left tail}
\eea
where
\bea
c_p= -\frac{p^2}{2 (p+1)(p+2)}\binom{p}{\frac{p}{2}}^{-\frac{2}{p}},
\eea
and $c=-\frac{1}{8}$ if $p=2$, $c=-\frac{1}{2}$ if $p>2$, and $C_p$ is a constant, possibly depending on $p$.
Although, the perturbative subleading corrections are not yet rigorously studied, but using the analogy to the $p=2$ case in Eq.\eqref{F2 exp corr}, we can conjecture,
\bea
o(1)= \log \left( 1 + b_1 |s|^{\frac{-2(p+1)}{p}} + \cO\left(|s|^{\frac{-4(p+1)}{p}}\right)\right),
\label{sublead pert mult}
\eea
for some undetermined constant $b_1$.
More explicitly, for $p=2,4,6,$ cases we have,
\begin{subequations}
\bea
\log F_2(s)= -\frac{|s|^3}{12} -\frac{1}{8} \log |s| + \log C_2 + o(1),
\eea
\bea
\log F_4(s)= -\frac{2}{45}\sqrt{6}\ |s|^{\frac{5}{2}} -\frac{1}{2} \log |s| + \log C_4 + o(1),
\eea
\bea
\log F_6(s)= -\frac{9}{560}20^{\frac{2}{3}}\ |s|^{\frac{7}{3}} -\frac{1}{2} \log |s| + \log C_6 + o(1).
\eea
\end{subequations}
Using the asymptotic \eqref{TWp left tail}, and the double scaling parameter \eqref{p-scaling}, $s=\lambda_p N^{\frac{p}{p+1}}$, we obtain the free energy in the strong coupling phase ($\beta>\beta_c$), as
\bea
\cF= \cF_c+ c_p |\lambda_p|^{\frac{2(p+1)}{p}} + N^{-2} \left(c \log |\lambda_p| + \frac{c\ p}{p+1} \log N + \log C_p\right) + o(N^{-2}).
\label{F multi strong}
\eea
Furthermore, in the phase $\beta>\beta_c$, and $(\beta-\beta_c)=\cO(1)$, we can re-arrange the above result, and obtain the genus expansion of the free energy of the generalized GWW model,
\bea
\cF= \tilde\cF_0+ \cF_{\text{pert}}^{(p)},
\eea
where the genus zero part and the higher genus subleading perturbative corrections are 
\begin{subequations}
\bea
\tilde\cF_0=\sum_{n=1}^\infty n\, t_n^2  +c_p |\lambda_p|^{\frac{2(p+1)}{p}},
\label{m-crit full genus zero free energy}
\eea
\bea
\cF_{\text{pert}}^{(p)}=\cF_1^{(p)}+  \sum_{g=2}^\infty N^{2-2g} \cF_g^{(p)},
\label{multi pert}
\eea
\end{subequations}
with the genus one free energy is given by
\bea
\cF_1^{(p)}= -\frac{1}{2}\log |\lambda_p|  -\frac{p}{2(p+1)} \log N + \log C_p.
\eea
For higher genus $g\geq 2$, by using the expansion of the conjecture, one can show that the higher than two genus free energy is given by \eqref{sublead pert mult},
\bea
o(1)\approx \sum_{n=1}^\infty (-1)^{n-1}\frac{b_1^n }{n} |s|^{-\frac{2(p+1)}{p}n}, \qquad \cF_g^{(p)}=\frac{(-1)^g}{g-1}\left(b_1\right)^{g-1} |\lambda_p|^{\frac{2(p+1)}{p}(1-g)}.
\eea
To summarize the results obtained in this Section for the free energy in the weak and strong phases of the multi-critical generalized GWW model, we have
\bea
    \cF^{(p)} = \sum_{n} n\, t_n^2+ \begin{cases}
    \displaystyle
        \cF_{\text{inst}}^{(p)} & \text{for } \beta<\beta_c\\[1em] \displaystyle
       c_p |\lambda_p|^{\frac{2(p+1)}{p}}+\cF_{\text{pert}}^{(p)} & \text{for } \beta>\beta_c
        \end{cases},
        \label{multi F}
\eea
where $\cF_{\text{inst}}^{(p)}$ and $\cF_{\text{pert}}^{(p)}$ are given by Eqs. \eqref{multi inst} and \eqref{multi pert}, respectively.

\subsubsection*{Cross-over region}
In this part, we discuss the smooth cross-over region between the weak and strong coupling phases, at large, but finite $N$, in the vicinity of the critical point $\beta_c$; $\beta>\beta_c$ and $(\beta-\beta_c)= \cO(N^{-\frac{p}{p+1}})$. The cross-over region is the intermediate region between the right and left tails, i.e. $\text{TW}^{(p)}(s)$ for finite $s\sim\alpha_p^{-\frac{1}{1+p}}$. The free energy of the generalized GWW multi-critical model in this region can be obtained from Eq.\eqref{TW-p}, as
\begin{align}
    \mathcal{F}=\sum_{n=1}^\infty n\, t_n^2 - N^{-2}\int_{s}^{\infty} (x-s)\ q^2\left((-1)^{\frac{p}{2}+1}x\right) \dd x
    \qquad \text{where} \qquad
    s\sim\alpha_p^{-\frac{1}{1+p}}= \cO(1),
\end{align}
and $q(s)$ is the solution of the $\frac{p}{2}$-th member of the Painlev\'{e} II hierarchy \cite{cafasso2019fredholm}, see also \cite{krajenbrink2020painleve},
\bea
\left(\dv{s}+2q\right)\ \cL_{p/2}[q_s-q^2]= s\, q,
\eea
where $q_s$ is the derivative of $q$ w.r.t. $s$, and operators $\cL_n$ are Lenard operators defined by 
\bea
\dv{s} \cL_{j+1} f = \left(\dv[3]{s}+4 f \dv{s}+ 2 f_s\right) \cL_j f, \quad \cL_0 f= \frac{1}{2}, \quad \cL_j 1 =0, \ j\geq 1.
\eea
Moreover, the multi-critical specific heat satisfies
\bea
q^2\left((-1)^{\frac{p}{2}+1}x\right)=-\dv[2]{s} \log F_p(s),
\eea
and in the strong coupling phase, as $s\to-\infty$, it has the following expansion \cite{cafasso2019fredholm},
\bea
q\left((-1)^{\frac{p}{2}+1} s\right) = \left(\frac{(\frac{p}{2})!^2}{(p)!}|s|\right)^{\frac{1}{p}}+ \frac{c}{2}\left(\frac{(p)!}{(\frac{p}{2})!^2}\right)^\frac{1}{p} |s|^{-2-\frac{1}{p}} + \cO\left(|s|^{-2-\frac{2}{p}}\right),
\eea
and, in the weak coupling phase, as $s\to\infty$, we have $q((-1)^{\frac{p}{2}+1} s)\to \text{Ai}_{p}(s)$. One can study the interpolation of $q(s)$ between these two limits and extract the finite $s$ results. 

\subsubsection{Multi-critical $1/N$ expansion}
Similar to the discussion in Section \ref{1/N expansion}, our results in this Section about the large $N$ dependence of the free energy of the multi-critical generalized GWW model in vicinity of the critical point, can be summarized in the following $1/N$ expansion,
\begin{align}
    \cF = \begin{cases}
    \displaystyle
        \cF_c+ \sum_{n=1}^\infty N^{-n-2}\ G_n^{\ (p,1)}\ e^{-2 N f_n^{(p)}(\beta)} & \text{for } \beta<\beta_c,  |\beta-\beta_
        c|= \cO(1) \\[1em] \displaystyle
        \cF_c +\sum_{n=1}^\infty N^{-\frac{p}{p+1}n-2}\ G_n^{\ (p,2)}  & \text{for } \beta<\beta_c, |\beta-\beta_
        c|= \cO(N^{-\frac{p}{p+1}})\\[1em] \displaystyle
        \cF_c+ \cF_0^{(p)} + N^{-2}\cG^{(p)}  -\sum_{n=1}^\infty N^{-2n-2}\ G_n^{\ (p,3)} & \text{for } \beta>\beta_c, |\beta-\beta_
        c|= \cO(1)
        \end{cases},
\end{align}
where $\cF_c=\sum_n n\, t_n^2$, and it is straightforward to observe that $G_n^{\ (p,1)}$  and $f_n^{(p)}(\beta)$ in the weak coupling phase are given by
\begin{subequations}\label{eq:G_coef}
\bea
G_n^{\ (p,1)}=\frac{1}{4^{n}n}\ \alpha_p^{\frac{n}{p}}\ (\beta_c-\beta)^{-\frac{(p+1)n}{p}},\qquad f_n^{(p)}(\beta)=\frac{n p}{p+1}\ \alpha_p^{-\frac{1}{p}}\ (\beta_c-\beta)^{\frac{p+1}{p}},
\label{G(p,1), f(p)}
\eea
and in the strong coupling phase, for the genus expansion of the free energy we have
\bea
\cF_0^{(p)} =c_p\ \alpha_p^{-\frac{2}{p}} |\beta_c-\beta|^{\frac{2(p+1)}{p}},
\label{F_0(p)}
\eea
\bea
\cG^{(p)}  = \log C_p +\frac{1}{2(1+p)}\log \alpha_p-\frac{p}{2(p+1)}\log N -\frac{1}{2} \log |\beta_c-\beta|,
\label{G1(p)}
\eea
and 
\bea
 G_n^{\ (p,3)}=\frac{(-1)^{n-1}\ b_1^n }{n}\ \alpha_p^{\frac{2n}{p}}\ |\beta_c-\beta|^{-\frac{2n(p+1)}{p}}.
 \label{G(p,3)}
\eea
\end{subequations}


\subsection{Infinite limit ($p\to\infty$) of multi-critical dynamics and bulk fluctuation}
\label{infinite limit}
Let us consider the generalized GWW model with the infinite degree polynomial potential and study the multi-critical dynamics at $p\to \infty$. In this limit the double scaling parameter~\eqref{p-scaling} becomes
\bea
s= e^{-1} (\beta_c-\beta)\, N,
\label{scalinh p Inf}
\eea
where we used the fact that the limit $p\to\infty$ in definition of $\alpha_p$ in Eq. \eqref{alpha beta p}, implies that only the term with  $n\sim p$ survives, and we have
\bea
\lim_{p\to\infty} \alpha_p^{-\frac{1}{1+p}} \sim \lim_{p\to\infty}\left(\frac{p^{p+1}}{p!}\right)^{-\frac{1}{1+p}}= e^{-1}.
\eea
It is interesting to compute the multi-critical $1/N$ expansion in this limit,
\begin{align}
    \cF = \begin{cases}
    \displaystyle
        \cF_c+ \sum_{n=1}^\infty N^{-n-2}\ G_n^{\ (\infty,1)}\ e^{-2 N f_n^{(\infty)}(\beta)} & \text{for } \beta<\beta_c, |\beta-\beta_
        c|= \cO(1) \\[1em] \displaystyle
        \cF_c +\sum_{n=1}^\infty N^{-n-2}\ G_n^{\ (\infty,2)}  & \text{for } \beta<\beta_c, |\beta-\beta_
        c|= \cO(N^{-1})\\[1em] \displaystyle
        \cF_c+ \cF_0^{(\infty)}  + N^{-2}\cG^{(\infty)}   -\sum_{n=1}^\infty N^{-2n-2}\ G_n^{\ (\infty,3)} & \text{for } \beta>\beta_c, |\beta-\beta_
        c|= \cO(1)
        \end{cases},
\end{align}
where one can directly compute the limit $p\to \infty$ of the functions in Eqs.\eqref{eq:G_coef}, as
\begin{subequations}
\bea
G_n^{\ (\infty,1)}=\frac{e^n }{4^{n}n}  (\beta_c-\beta)^{-n},\qquad f_n^{(\infty)}(\beta)=\frac{n}{e}\  (\beta_c-\beta),
\eea
\bea
\cF_0^{(\infty)}  =-\frac{1}{8 e^2}  |\beta_c-\beta|^2, \qquad
\cG^{(\infty)}  = \log C_p +\frac{1}{2}-\frac{1}{2}\log N -\frac{1}{2} \log |\beta_c-\beta| ,
\eea
\bea
 G_n^{\ (\infty,3)}=\frac{(-1)^{n-1}\ b_1^n\ e^{2n} }{n} |\beta_c-\beta|^{-2n}.
\eea
\end{subequations}
In this limit, the free energy expansion in the strong coupling phase $\beta>\beta_c$, becomes
\bea
\cF^{(\infty)}(\beta)=\cF_c -\frac{1}{8 e^2} |\beta_c-\beta|^2 - \frac{1}{2N^{-2}} \left(\log |\beta_c-\beta|+ \log N-1\right) + \cO(N^{-4}),
\eea
and the rest of the subleading terms are
\bea
\sum_{n=1}^\infty \frac{(-1)^{n-1}\ b_1^n\, e^{2n}}{n} |\beta_c-\beta|^{-2n} N^{-2n-2}. 
\eea
The leading contribution to free energy indicates a second order phase transition. Moreover, the higher genus ($g\geq 2$) free energy  becomes
\bea
\cF_g^{(\infty)} (\beta)=\frac{(-1)^g\ \left(e^2\,b_1\right)^{g-1}}{g-1} |\beta_c-\beta|^{2-2g}.
\eea

In the weak coupling phase $\beta< \beta_c$, using the asymptotic behavior of the higher Airy function in Eq.\eqref{higher Airy asymp},
\bea
\lim_{p\to\infty} \text{Ai}_{p}(s)=a_\infty s^{-\frac{1}{2}}\ e^{-s}\left(1-\frac{\tilde{a}_\infty}{s}+ \cO\qty(\frac{1}{s^2})\right),
\eea
and the infinite limit of the $\text{TW}^{(p)}$ in Eq.\eqref{higher TW asymp 1} is 
\bea
\lim_{p\to\infty}F_p(s)\approx 1-\frac{e^{-2 s}}{2} +s\ E_1(2s)=1-\frac{e^{-2 s}}{2} +s\ \Gamma(0,2s),
\eea
where $\Gamma(t,x)$ is the upper incomplete Gamma function defined as
\bea
\Gamma(t,x)=\int_{x}^{\infty} s^{t-1} e^{-s}\, \dd s, 
\eea
and we used $\Gamma(t,x)=x^t\ E_{1-t}\left(x\right)$.
Then, we can expand the exponential integral or the Gamma function to obtain
\bea
\lim_{p\to\infty}F_p(s)\approx 1-\frac{1}{4} s^{-1}e^{-2s} \left(1-2s^{-1} 
+\cO(s^{-2})\right),
\eea
and then after keeping the first leading terms, we obtain
\begin{align}
    \lim_{p\to\infty}\log F_p(s) \approx
    \log \left( 1 - 4^{-1} s^{-1}e^{-2s} + \cO(e^{-2s}s^{-2}) \right)=-\sum_{n=1}^\infty\frac{1 }{4^n n}s^{-n}e^{-2n s}+ \cO(s\log s).
\end{align}
Finally, the infinite limit of the multi-critical instanton free energy can be obtained from Eq.\eqref{multi inst}, as
\bea
\cF^{(\infty)}_{\text{inst}}(\beta)= -\sum_{n=1}^\infty\frac{e^n}{4^n n}(\beta_c-\beta)^{-n} N^{-n-2}\exp{-2n e^{-1}(\beta_c-\beta) N},
\eea
and the one-instanton sector is
\bea
\cF^{(
\infty)}_{1\text{-inst}}(\beta)=- 4^{-1}e (\beta_c-\beta)^{-1} N^{-3}\exp{-2 e^{-1}(\beta_c-\beta) N}.
\eea

\subsection{Multi-critical phase structure and its interpretation}
Based on our analysis in this Section, we observe that in the generalized GWW model there is a multi-critical phase transition at critical point $\beta_c=\gamma$, meaning that the free energy has different expansion at left and right side of the critical point,
\bea
    \cF^{(p)} = \begin{cases}
    \displaystyle
        \cF_c+\cO(e^{-cN}) & \text{for } \beta<\beta_c\\[1em] \displaystyle
      \cF_c+c_p\ |\lambda_p|^{\frac{2(p+1)}{p}}+\cO(N^{-2}) & \text{for } \beta>\beta_c
        \end{cases}.
\eea
where $\cF_c= \sum_{n} n\, t_n^2$, 
\bea
c_p= -\frac{p^2}{2 (p+1)(p+2)}\binom{p}{\frac{p}{2}}^{-\frac{2}{p}},\qquad \text{and} \quad \lambda_p=\alpha_p^{-\frac{1}{1+p}} (\beta_c- \beta).
\eea
The leading order of the free energy in two phases, imply that the order of the multi-critical phase transition is $\frac{2(p+1)}{p}$. Often, in the literature this multi-critical phase transition, for any finite $p$, is considered a transition of order $\lfloor\frac{2(p+1)}{p}\rfloor = 3$. 
\subsubsection{Multi-critical gap dynamics}
\label{Multi-critical gap dynamics}
Before discussing the interpretations of the multi-critical phase transition, let us briefly explain some related issues. The assumption $\alpha_{p'}=0$ for $p'<p$, which is used to derive the coefficients \eqref{alpha beta p} in Eq.\eqref{TWp}, implies some physical constraints on the dynamics of the model. Notice that at any fixed $p$, the interaction and fluctuation are at scale $N^{-\frac{p}{p+1}}$, and thus in order to see the sub-dominant fluctuations for higher $p$, all the larger scale interactions at $p'<p$, should be turned off as we want to isolate the interaction at the scale $p$, and  study the effect of a fixed scale fluctuation. In general, to study the multi-critical dynamics, one needs the fine tuning of the coupling, for example in this case, introducing $p$-dependent, at any $p\geq 2$, couplings in the large $N$ limit, such as $ t_n^{(p)}\sim q^{(p-p')N}$, for $q<1$ and $2\leq p'\leq p$.

In the critical dynamics, we studied the third-order phase transition associated with one-gap dynamics. More precisely, in the generalized GWW model with the maximum degree $m$ polynomial potential,
\bea
V(U)= \sum_{n=1}^m t_n \tr(U^{n}+U^{-n}),
\eea
the critical dynamics is causing the transition between $l$-cut solution and $(l-1)$-cut solution for $1\leq l\leq m$.
In the multi-critical dynamics we must consider the multi-gap dynamics. The unitary matrix model with degree $m$ polynomial potential can have $1$- to $m$-cut solutions and could possibly have phase transitions between $l$-cut solution to $(l-r)$-cut solution for $0\leq r\leq l\leq m$. These phase transitions are associated with the colliding of the two or more end points of the cuts, simultaneously at the same critical point $\beta_c$, which depends only on the potential. Let us isolate a particular set of phase transitions, namely the transitions between the $\frac{p}{2}$-cut solutions to full support (zero gap) solution, in which $p$-end points of the cuts collide simultaneously, for $p\leq 2m$. We call this transition a $p$-multi-critical dynamics for $p>2$. The case $p=2$ is called critical dynamics. There are $m$ possible phase transitions in this set, labeled by $2 \leq p\leq 2m$, and each different $p$-(multi)-critical dynamics is governing the opening/closing of $\frac{p}{2}$-gaps as a result of the edge fluctuation of order $N^{-\frac{p}{p+1}}$ at critical point. In our study, there is no mixing between the dynamics and fluctuations at different $p$, as we isolate the fluctuation scale $N^{-\frac{p}{p+1}}$ by putting $\alpha_{p'}=0$, for $p'<p$. We did compute the free energy of the generalized GWW model near the critical point $\beta_c$ with the $p$-end points of the cuts are colliding simultaneously.



\subsubsection{A new second-order phase transition}
In the limit $p\to\infty$, the asymptotic behavior of the Fredholm determinant \eqref{TWp left tail} is
\bea
\lim_{p\to\infty} F_p(s)\sim |s|^{-\frac{1}{2}} e^{-\frac{1}{8} |s|^2},
\label{inf FD}
\eea
where the scaling parameter is $s\sim (\beta_c-\beta) N$.
This is similar to the asymptotics of the Fredholm determinant with the sine kernel discussed in Section~\ref{bulk fluc}, and in fact indicating of the bulk fluctuation. 
From the physical point of view, in the infinite limit, there are infinite number of the gaps as well as infinite number of the cuts. Then, the infinite-multi-critical dynamics is related to the shrinking of the infinite gaps at a critical point, when infinite end points of the cuts collide, simultaneously. The infinite number of the gaps can be seen as the phase of the full gap, i.e. no support for eigenvalues. Thus, when all the gaps shrink simultaneously, the model undergoes a transition to the phase of full support for the eigenvalues or in other words, the infinite number of cuts collide and eigenvalues cover the full circle. 
As we observe from Eq.\eqref{inf FD}, the phase transition between the full support and full gap is of the second order. This is a phase transition associated with the infinite number of edge fluctuations or in other words a zero-to-full support transition driven by the bulk fluctuation.

\subsection*{Higher Plancherel partition and multi-critical models}
In this part, we consider some basic examples of the unitary matrix models with multi-critical dynamics. 
First, consider the degree two polynomial potential with the couplings,
\bea
t_n= t_1\delta_{n,1}+t_2\delta_{n,2}.
\eea
In this model, the multi-critical parameter is restricted to $p\leq 4$, and the parameters can be easily computed
\begin{align}
    \alpha_p =\frac{2}{p!} (t_1+ 2^{1+p} t_2) 
    \, , \quad
    \beta = 2 (t_1+2 t_2) 
     \, , \quad
     \lambda_p= 2^{-\frac{1}{1+p}}(1-2t_1-4t_2)\left(\frac{t_1+2^{1+p}t_2}{p!}\right)^{-\frac{1}{1+p}}.
    \label{}
\end{align}
The critical point $\beta_c=1$ implies that the surface of critical couplings is $2t_1^*+4t_2^* -1 =0$.

Using the above parameters, we can compute the first few terms of the free energy in the strong coupling regime can be computed from Eq.\eqref{F multi strong}, as
\bea
\cF\approx t_1^2+2t_2^2+c_p |\lambda_p|^{\frac{2(p+1)}{p}}+ \frac{c}{N^2}\log |\lambda_p N^{\frac{p}{p+1}}|.
\eea
The higher order corrections to the free energy can be computed in a straightforward way.
Let us consider the critical and multi-critical cases separately. In the critical case $p=2$, the double-scaling parameter is
\begin{align}
    \lambda_2 =  \frac{1-2t_1-4t_2}{(t_1+8 t_2)^{1/3}},
    \label{}
\end{align}
and the first few terms of the free energy are obtained as
\bea
 \cF \approx t_1^2+2t_2^2-\frac{(-1+2t_1+4 t_2)^3}{12 (t_1+8 t_2)}-\frac{1}{8 N^2}
    \log \left|\frac{1-2t_1-4t_2}{(t_1+8 t_2)^{1/3}} \right| -\frac{\log N}{12 N^2}.
\eea
In the multi-critical case $p=4$, in principle we can fine tune the couplings $t_1$, and $t_2$ such that $\alpha_2 \to 0$, and similarly one can compute,
\begin{align}
    \lambda_4 =\frac{2^{2/5}\, 3^{1/5}\,(1-2t_1-4t_2)}{(t_1+32 t_2)^{1/5}},
    \label{}
\end{align}
and the leading orders in free energy as
\bea
\cF \approx t_1^2+2t_2^2+ \frac{4}{15}\sqrt{2}\left(\frac{1-2t_1-4t_2}{(t_1+32 t_2)^{1/5}}{}\right)^{5/2}-\frac{1}{2N^2}
    \log \left|\frac{1-2t_1-4t_2}{(t_1+32 t_2)^{1/5}} \right| -\frac{2\log N}{5 N^2}.
\eea
Moreover, the above results can be expanded around the critical surface, for example in terms of the critical coupling $t_1^*$, by using $t_2^*= (1-2t_1^*)/4$.

\section{Supersymmetric indices: Examples}\label{example}
In this Section, we apply the machinery of the generalized GWW model to study the matrix integral representation of the indices of the gauge theory in the large $N$ limit. Although the exact relation is provided through Hubbard--Stratonovich transformation, we consider this approximation as a toy model for gauge theory indices, and use the same couplings, from generalized GWW model, in the potential of the matrix integrals of the gauge theories. Using this approximation, we study the perturbative and non-perturbative regimes of gauge theory and explicitly compute the free energy and phase structure, in some concrete examples with some interesting choices, motivated by gauge theory, of finely tuned couplings such that the matrix model is solvable. The results in this part are obtained from the direct computations of the contributions of the left and right tails of $\text{TW}^{(p)}$, to the free energy in Section \ref{multi section}. 

\subsection{Hagedorn phase transition and deconfinement}
Before, diving to the explicit computations in some concrete examples, let us discuss the Hagedorn phase transition in gauge theories and its relation to the deconfinement transition in the generalized GWW model, discussed in this paper, so far.
The Hagedorn transition is the phase transition, in the large $N$ limit, between the domain of the convergence and divergence in the generating function, separated by the hyper-surface of the singularities of the generating functions, corresponding to Hagedorn phase transition point.
The matrix integral of the gauge theory indices $\cI$ in Eq.\eqref{index PE}, in the large $N$ limit, using the Gaussian matrix integration, becomes an infinite product \cite{Aharony:2003sx},
\bea
\lim_{N\rightarrow \infty}\cI(q_i)= \prod_{n=1}^\infty \frac{1}{1-f(q_i^n)}.
\eea
Thus, using the above generating function at large $N$ limit,
we can study the singularities of the generating function at the poles of the infinite product formula. First, consider $n=1$ which corresponds to the double-trace matrix model and can effectively be approximated by GWW model. Notice that in this case, the Hagedorn critical point and GWW critical point coincide $t^*=f(q_i^*) =1$.
In the general case, we observe that the critical hyper-surface of the Hagedorn transition is given by 
\bea
\prod_{n=1}^\infty \left(1-f(q_i^{*n})\right) = 1- \sum_{n=1}^\infty f(q_i^{*n}) + \cO(f^2(q_i^{*n}))= 0.
\label{Hag crit hyp}
\eea

On the other hand, in the generalized GWW model, the parameters $\alpha$ and $\beta$ depend on the couplings $t_n$, and the couplings can be any univariate or multivariate function of the parameters of the gauge theory $q_i$, thus we have
\begin{align}
    \alpha_p (q_i) = 2\sum_{n=1}^\infty \frac{n^{p}}{p!} \, f(q_i^n)
    \, , \qquad
    \beta (q_i) = 2\sum_{n=1}^\infty  f(q_i^n)
    \, .
    \label{}
\end{align}
The strong-weak coupling or the deconfinement phase transition in this model, which is a generalization of the Gross--Witten phase transition, happens at $\beta_c (q_i)=\gamma=1$, between the two phases of weak-coupling $\beta (q_i)<1$, and strong-coupling $ \beta (q_i)>1$. 

Using the couplings $t_n= f(q_i^n)/n$, and re-scaling all of them by a factor $1/2$, similarly to what we have in the original GWW model, the critical hyper-surface of deconfinement transition in the generalized GWW model is given by
\bea
\beta_c=\sum_{n=1}^\infty n\, t^*_n =\sum_{n=1}^\infty f(q_i^{*n})= 1.
\label{deconf crit hyp}
\eea
Then, by expanding the Hagedorn critical hyper-surface \eqref{Hag crit hyp} and comparing it with deconfinement critical hyper-surface \eqref{deconf crit hyp}, we observe that they match up to linear order in $f(q_i)$. The single letter indices $f(q_i)$ in the domain of convergence, $f(q_i)<1$, approaches the Hagedorn critical hyper-surface, $f(q_i)\nearrow 1$,
and thus the deconfinement transition in generalized GWW model effectively explains the Hagedorn phase transition.

In summary, we observe that the generalized GWW model and generalized double-trace model have qualitatively similar phase structure. In fact, upon identification of the couplings in two models, the phase structure of generalized GWW model captures the dominant contribution, i.e. linear approximation in Hagedorn phase structure. This implies a universal deconfinement/Hagedorn phase transition at critical point $\beta_c=1$, for any arbitrary coefficient $f(q_i)$ of the potential in the matrix integral of the gauge theory.
This is the phase transition associated with the gap opening in the multi-critical matrix model.

A priori, the gauge theory double-trace matrix model is approximated by the single-trace GWW model in the large $N$ limit, thus we expect that the results of GWW model carry over to the double-trace model, at least up to leading order and possibly extending to first few subleading corrections. A possible clue for roughly examining that to what extent the subleading corrections of single-trace model apply to double-trace model, comes from the observation made in this section, that the critical point of the single-trace and double trace models coincide up to linear order in $f(q_i^{*n})$, see Eq.\eqref{deconf crit hyp}. At $n=1$ case, the critical temperature of the GWW model exactly matches with that of double-trace model. This indicates that results for $n=1$ case, can more accurately transfer to the double-trace model. Moreover, we can expect that in the large $N$ limit, all the results about the free energy and phase structure can safely carry over up to linear order in $f(q_i^{*n})$.  More precise statement about the subleading corrections at finite $N$ requires an exact analysis of the Hubbard-Stratonovich transformation but we expect that the Legendre transform of the current leading and subleading results would apply to gauge theory indices.
Finally, notice that in this paper, we do not directly study the double-trace matrix model, and  only consider the toy models which are generalized GWW models with the coefficients in the potential obtained from the double-trace models.







\subsection*{Trivial example}
Before studying some physically motivated examples, we consider an interesting mathematical example with the fine-tuned couplings $t_n^{(p)}=q^{(p-p')N}/n^s$ which, in the limit $N \to \infty$, is zero for $2\leq p'< p$, and $t_n=1/n^s$, for $p'=p$. This example corresponds to the matrix model with the following poly-logarithmic potential
\bea
V(U) = \sum_n t_n (\tr U^n+\tr U^{-n})= \left(\text{Li}_s(e^{\ii \theta})+\text{Li}_s(e^{-\ii \theta})\right). 
\eea
It is easy to compute the parameters of this model in terms of the Riemann zeta function,
\begin{align}
    \alpha_p = \frac{2}{p!}\sum_{n=1}^\infty n^{p+1} \, t_n = \frac{2}{p!} \zeta(s-p-1)
    \, , \qquad
    \beta = 2\sum_{n=1}^\infty n\, t_n=2\zeta(s-1)
    \, .
    \label{}
\end{align}
However, as the parameter $\beta$ is a coupling independent number, there is no phase transition in this model. The free energy of the model is the continuum free energy
\bea
\cF_c=\sum_n n\, t_n^2= \zeta(2s-1).
\eea
In particular, the especial case of $s=1$ is a non-interacting theory with zero potential, with $\alpha_p=0$ for all $p\in 2\mathbb{N}$. Moreover, the free energy is divergent at $s=1$ as $\zeta(1)\to \infty$. 
Next, we consider a model with an arbitrary degree single term potential,
\bea
t_n=t_k^{(p)}\delta_{n,k}, \qquad \text{with}\qquad t_k^{(p)}=\frac{t^k}{k}\, q^{(p-p')N},
\eea
where $q<1$, and for all $p'<p$, we have $t_k=0$, at large $N$ limit.
In this model, after fine-tuning the coupling for any given $p$ such that $\alpha_{p'}=0$, for $p'<p$, the parameters of the model can be computed as
\begin{align}
    \alpha_p = \frac{2k^p}{p!} t^k 
    \, , \qquad
    \beta = 2 t^k
    \, , \qquad
    \lambda_p=2^{-\frac{1}{1+p}}(1-2t^k)\left(\frac{k^p t^k}{p!}\right)^{-\frac{1}{1+p}},
    \label{}
\end{align}
and the genus zero free energy in the strong coupling regime is
\bea
 \tilde\cF_0^{(p)}=\frac{t^{2k}}{k}+2^{-\frac{2}{p}-1}\frac{p^2}{ (p+1)(p+2)}\binom{p}{\frac{p}{2}}^{-\frac{2}{p}}\left(\frac{k^p t^k}{p!}\right)^{-\frac{2}{p}} \left(1-2t^k\right)^{\frac{2(p+1)}{p}}.
\eea
Since $p\leq 2k$, the infinite multi-critical limit, $p\to\infty$, implies $k\to \infty$ and one can assume $p=2k$. In this limit although $\alpha_2$ and $\beta$ are either zero or diverging depending on $|t|$, but $\lambda_\infty$ is finite and one can show $\lambda_\infty= \frac{2}{e\sqrt{t}}$ and thus the free energy becomes
\bea
\tilde\cF_0^{(\infty)}=\frac{t^{2k}}{k} -\frac{t^{-1}}{2 e^2}.
\eea

\subsection{Free chiral ring index}
\begin{figure}
\centering
\begin{tikzcd}
\bullet \arrow[out=60,in=120,loop,swap,"t_1"]
\end{tikzcd}
\begin{tikzcd}
\bullet \arrow[out=0,in=60,loop,swap,"t_1"]
  \arrow[out=120,in=180,loop,swap,"t_2"]
  \arrow[out=240,in=300,loop,swap,"t_3"]
  \end{tikzcd}
\begin{tikzcd}
\bullet \arrow[out=0,in=60,loop,swap,"t_1"]
  \arrow[out=90,in=150,loop,swap,"t_2"]
  \arrow[out=180,in=240,loop,swap,"\cdots"]
  \arrow[out=270,in=330,loop,swap,"t_k"]
\end{tikzcd}
\caption{From left to right: Jordan quiver, clover quiver, and generalized clover quiver}
\label{clover}
\end{figure}
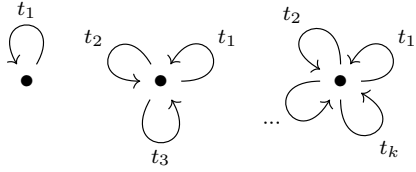
A class of gauge invariant operators which are annihilated by all the supercharges of one chirality, for example positive R-charge, are called chiral operators. The local gauge invariant chiral operators form a commutative ring, called chiral ring. 
In this Section we consider the matrix integral representation of the generating function for the counting of the multi-trace chiral gauge invariant operators in the chiral ring of a generic free $\cN=1$ superconformal quiver theories \cite{pasukonis2013quivers}.
Quiver gauge theory is a gauge theory encoded on a graph, consists of nodes and arrows representing gauge groups and chiral fields, respectively. The unitary gauge group $G$ of the quiver is a product over the $\mathrm{U}(N)$ factors of each node $i$ of quiver, $G= \prod_i \mathrm{U}(N_i)$.
The generating function of the BPS operators in the free chiral ring is represented by the following quiver multi-matrix integral,
\bea
\mathcal{I}(q_i) = \int_{\mathrm{U}(N)} \prod_a \dd{U_a} \exp\left[\sum_{n=1}^\infty\sum_{a,b,\alpha} \frac{1}{n}\left(q_{ab;\alpha}\right)^n \tr U_a^n \tr U_b^{-n}\right],
\label{quiver MI}
\eea
where the $q_{ab;\alpha}$ are the fugacity factors associated with the chiral fields $\{\phi_{ab;\alpha}: \alpha\in \{1, 2, ..., K_{ab}\}\}$ with $K_{ab}$ number of arrows between nodes $a$ and $b$, and the chiral fields are transforming in the bifundamental representation $(N_a, \bar N_b)$. The chiral fields in the adjoint representation of a gauge group are the loops starting and ending on the same node of the quiver associated to the gauge group.

In the large $N$ limit, one can derive the generating function as an infinite product, 
\bea
\lim_{N\rightarrow \infty}\cI(q_i)= \prod_{n=1}^\infty \frac{1}{\det\left[1-\cA(q_i^n)\right]},
\eea
where $\cA(q_1, q_2, ...)$ is the adjacency matrix of the quiver, equipped with the fugacity-weighted arrows. 

The simplest physical case is the clover quiver theory, i.e. $\cN=4$ super Yang-Mills theory, a quiver with one node associated with the $\mathrm{U}(N)$ gauge group, and three arrows or the chiral fields in the adjoint representation of $\mathrm{U}(N)$. Trivially, it can be generalized to the case of the $k$ chiral fields, see Fig.~\ref{clover}, which is often called in the literature by the $k$-matrix model harmonic oscillator. For example, the case $k=2$, the 2-matrix harmonic oscillator, is considered in \cite{Aharony:2003sx}. The $k=3$ case is $\cN=4$ SYM and $k=4$ case is the conifold theory. The asymptotics and phase structure of the generalized clover quiver with $k$ chiral fields, are studied in \cite{ramgoolam2020quiver}.
The generating function of the index of the generalized clover quiver can be obtained from Eq.\eqref{quiver MI}, as the following one-matrix model,
\bea
\label{clover MI}
\mathcal{I}(q_i) = \int_{\mathrm{U}(N)} \hspace{-1em} \dd{U} \exp\left[\sum_{n=1}^\infty \frac{1}{n}(q_1^n+ q_2^n + ...+ q_k^n)  \tr U^n \tr U^{-n}\right].
\eea
This matrix model can be approximated in the large $N$ limit by the generalized GWW model with coupling $t_n= \frac{1}{n}\sum_j q_j^n$.
In this approximation, this matrix model can be seen as a unitary matrix model with the following potential, 
\bea
V(U)=\tr\log \left(\prod_{j=1}^k (1-q_j U) (1-q_j U^{-1})\right).
\eea

\subsubsection*{Jordan quiver}
Let us first consider the univariate case of the generalized clover $q_j=q_1 \delta_{j,1}$, and change variable $q_1=t$ for more convenience. This is a case of clover quiver with one arrow, and it is known as the Jordan quiver. This is also an example of the simplest matrix model with plethystic exponential potential. Similarly, the Jordan quiver gauge theory can be approximated by the generalized GWW matrix model with the coupling $t_n= t^n/n$, which leads to the matrix model with the following potential,
\bea
V(U)=\tr\log (1-t U) (1-t U^{-1}).
\eea

In this form of the coupling ($t_n= t^n/n$), as the assumption $\alpha_{p'}=0$ for $p'<p$, can not be satisfied, there is only a possibility for the $p=2$ critical dynamics, and in fact, we can only discuss the multi-critical dynamics once we fine-tune the couplings, similar to what we have done in the previous parts. For the fine-tuned couplings, one can find the following generating functions for the parameters $\alpha_p$ and $\beta$, 
\begin{align}
    \alpha_p = \frac{2}{p!}\sum_{n=1}^\infty n^{p} \, t^n = \frac{2}{p!} \text{Li}_{-p}(t)
    \, , \qquad
    \beta = 2\sum_{n=1}^\infty t^n=\text{Li}_{0}(t)= \frac{2t}{1-t}
    \, ,
    \label{}
\end{align}
and the parameter $\lambda_p$ is computed as 
\begin{align}
    \lambda_p =\alpha_p^{-\frac{1}{1+p}} (\beta_c- \beta)= 2^{-\frac{1}{1+p}}\left(\frac{1-3t}{1-t}\right)\left(\frac{\text{Li}_{-p}(t)}{p!}\right)^{-\frac{1}{1+p}}.
    \label{}
\end{align}
First notice that critical temperature defined by $\mu_c= -\log t_c$, can be obtained from $\beta_c=1$, ($t_c=1/3$), as $\mu_c = \log 3$.
Then, we can compute the free energy in the weak and strong coupling phases from Eq.\eqref{multi F}, as
\bea
\cF=\begin{cases}
    \displaystyle
        -\log(1-t^2)+\cF^{(p)}_{\text{inst}} & \text{for } t<1/3\\[1em] \displaystyle
       -\log(1-t^2)+2^{-\frac{2}{p}}c_p \left(\frac{1-3t}{1-t}\right)^{\frac{2(p+1)}{p}}\left(\frac{\text{Li}_{-p}(t)}{p!}\right)^{-\frac{2}{p}}+
       \cF^{(p)}_{\text{pert}} & \text{for } t>1/3
        \end{cases},\nonumber\\  
        \label{Jordan F}
\eea
where $\cF^{(p)}_{\text{inst}}$ is computed by sum over all $n$-instantons contributions, given by 
\bea
\cF^{(p)}_{n\text{-inst}}=\frac{1 }{4^{n}n}\,\xi_p^{-n} N^{-n-2}\exp{-\frac{2n p}{p+1}\xi_p N},
\eea
where 
\bea
\xi_p=\lambda_p^{\frac{p+1}{p}}= 2^{-\frac{1}{p}}\left(\frac{1-3t}{1-t}\right)^{\frac{p+1}{p}}\left(\frac{\text{Li}_{-p}(t)}{p!}\right)^{-\frac{1}{p}},
\eea
and $\cF^{(p)}_{\text{pert}}$ is given by
\bea
\cF_1^{(p)}= -\frac{c}{1+p}\log 2+ c \log \left(\frac{1-3t}{1-t}\right)-\frac{c}{1+p} \log \left(\frac{\text{Li}_{-p}(t)}{p!}\right) -\frac{p}{2(p+1)} \log N + \log C_p,\nonumber\\
\eea
and
\bea
\cF_g^{(p)}=\frac{(-1)^g\, (c_1)^{g-1} }{2^{\frac{2-2g}{p}}(g-1)} \left(\frac{1-3t}{1-t}\right)^{\frac{(p+1)}{p}(2-2g)}\left(\frac{\text{Li}_{-p}(t)}{p!}\right)^{-\frac{2-2g}{p}}.
\eea
The leading genus zero free energy \eqref{m-crit full genus zero free energy} can be obtained as
\begin{align}
    \tilde\cF_0=-\log(1-t^2)-2^{-\frac{2}{p}-1}\frac{p^2}{ (p+1)(p+2)}\binom{p}{\frac{p}{2}}^{-\frac{2}{p}} \left(\frac{1-3t}{1-t}\right)^{\frac{2(p+1)}{p}}\left(\frac{\text{Li}_{-p}(t)}{p!}\right)^{-\frac{2}{p}}.
\end{align}
Let us consider the critical case $p=2$, in which we have
\begin{align}
    \alpha_2 = \sum_{n=1}^\infty n^{2} \, t^n =\text{Li}_{-2}(t)= \frac{t(1+t)}{(1-t)^3}
    \, , \qquad
    \beta = \frac{2t}{1-t}
    \, ,
    \label{}
\end{align}
Similar results for this matrix model are obtained, from Riemann--Hilbert method, in \cite{baik2000random, baik2001symmetrized}. 
The double-scaling parameter $s$ and the genus-zero free energy in this case are obtained as 
\begin{align}
    \lambda_2 = \frac{1-3t}{t^{\frac{1}{3}}(1+t)^{\frac{1}{3}}}
    \, , \qquad
     \tilde\cF_0=-\log(1-t^2)-\frac{(1-3t)^3}{12t(1+t)}.
    \label{}
\end{align}
As an example of multi-critical dynamics for the model with the fine tuned couplings, let us consider the case $p=4$, in which we have
\begin{align}
    \alpha_4 =\text{Li}_{-4}(t)= \frac{t(1+t)(1+t(10+t))}{12(1-t)^5}
    \, , \qquad
    \beta = \frac{2t}{1-t}
    \, .
    \label{}
\end{align}
The parameter $\lambda_4$ and the genus zero free energy are obtained as 
\begin{align}
    \lambda_4 = \frac{12^{1/5} (1-3t)}{\left(t(1+t)(1+t(10+t))\right)^{1/5}}
    \, , \quad
     \tilde\cF_0=-\log(1-t^2)+\frac{4}{15}\sqrt{2}\frac{(1-3t)^{5/2}}{\left(t(1+t)(1+t(10+t))\right)^2}.
    \label{}
\end{align}

It is also interesting to consider the limit $p\to \infty$, and by using we observe
\bea
\lim_{p\to\infty} \text{Li}_{-p}(e^{-\mu})= \Gamma(1+p) \mu^{-p-1},\quad \lim_{p\to\infty} \left(\frac{\Gamma(1+p) }{p!}\right)^{-\frac{1}{1+p}}=1,
\eea
which lead to
\bea
\lambda_\infty=\xi_\infty= \frac{1-3t}{1-t}\mu,
\eea
and the free energy,
\bea
    \cF^{(\infty)} =  -\log(1-t^2)+ \begin{cases}
    \displaystyle
        \cF_{\text{inst}}^{(\infty)} & \text{for } t<1/3\\[1em] \displaystyle
      \frac{1}{16} \left|\frac{1-3t}{1-t}\mu\right|^{2}+\cF_{\text{pert}}^{(\infty)} & \text{for } t>1/3
        \end{cases},
\eea
where $\cF_{\text{inst}}^{(\infty)}$ is the sum over
\bea
\cF^{(\infty)}_{n-\text{inst}}=\frac{1}{4^{n} n}\left(\frac{1-3t}{1-t}\mu\right)^{-n} N^{-n-2}\exp{2\left(\frac{1-3t}{1-t}\right) \mu nN},
\eea
and $\cF^{(\infty)}_{\text{pert}}$ is given by
\bea
\cF_1^{(\infty)}= c \log \left(\frac{1-3t}{1-t}\right) -\frac{1}{2} \log N + \log C_p,\quad
\text{and} \quad
\cF_g^{(\infty)}=\frac{(-1)^g }{g-1}c_1^{g-1} \left(\frac{1-3t}{1-t}\mu\right)^{2-2g}.\nonumber\\
\eea
\subsubsection*{Generalized clover quiver}

Using the large $N$ limit of the generating function of the generalized clover quiver,
\bea
\lim_{N\rightarrow \infty}\cI(q_i)= \prod_{n=1}^\infty \frac{1}{1-\sum_{j=1}^k q_j^n}.
\eea
As it is shown in \cite{ramgoolam2020quiver}, the leading singularity of the above generating function is at $n=1$, leading to the Hagedorn phase structure of this model with the critical hyper-surface  $\sum_{j=1}^k q^*_j=1$. Thus, in our approximation, the GWW model is dominant partition function of this model, with the parameters at the critical dynamics $p=2$, 
\bea
\alpha_2= \sum_{j=1}^k q_j\equiv T, \qquad \beta= 2\sum_{j=1}^k q_j=2T.
\eea
Similarly, all the results in Section \ref{GWW section} follow immediately by replacing $t_1$ with $T$.
Moreover, straightforward generalization of the results for Jordan quiver to the generalized clover quiver leads to
\begin{align}
    \alpha_p =  \frac{2}{p!}\sum_{j=1}^k \text{Li}_{-p}(q_j)
    \, , \qquad
    \beta = \sum_{j=1}^k \frac{2q_j}{1-q_j}
    \, ,
    \label{}
\end{align}
and the parameter $\lambda_p$ is obtained as 
\begin{align}
    \lambda_p = 2^{-\frac{1}{1+p}}(p!)^{\frac{1}{1+p}}\left(\sum_{j=1}^k \text{Li}_{-p}(q_j)\right)^{-\frac{1}{1+p}}\left(1-\sum_{j=1}^k \frac{2q_j}{1-q_j}\right).
    \label{gen clov lambda}
\end{align}
Then, it is straightforward to compute $\cF^{(p)}$, $\cF^{(p)}_{\text{inst}}$ and $\cF^{(p)}_{\text{pert}}$ by inserting $\lambda_p$ from \eqref{gen clov lambda} in Eqs.\eqref{multi F}, \eqref{multi inst} and \eqref{multi pert}.
For example, the genus zero energy can be obtained as
\bea
\tilde\cF_0=\sum_{n} \frac{1}{n}\, \left(\sum_{j=1}^k q_j^n\right)^2+ 2^{-\frac{2}{p}-1}\frac{p^2 p!^{\frac{2}{p}}}{ (p+1)(p+2)}\binom{p}{\frac{p}{2}}^{-\frac{2}{p}} \left(\sum_{j=1}^k \text{Li}_{-p}(q_j)\right)^{-\frac{2}{p}}\left(1-\sum_{j=1}^k \frac{2q_j}{1-q_j}\right)^{\frac{2(p+1)}{p}}.\nonumber\\
\eea
Moreover, in the limit $p\to\infty$, the smallest $\mu_j=-\log q_j$, denoted by $\mu_{\text{min}}$, contributes in $\lambda_\infty$, as we obtain
\begin{align}
    \lambda_\infty = \left(1-\sum_{j=1}^k \frac{2q_j}{1-q_j}\right)\mu_{\text{min}},
    \label{}
\end{align}
and, using this parameter, the free energy can be computed in a similar fashion.

Finally, in the unrefined case $q_1= q_2 = ...=q_k\equiv q$, one can compute the parameters 
\begin{align}
    \alpha_p =  \frac{2k}{p!} \text{Li}_{-p}(q)
    \, , \qquad
    \beta = \frac{2k q}{1-q}
    \, , \qquad
    q_c = \frac{1}{2k+1},
    \label{}
\end{align}
and
\begin{align}
    \lambda_p = (2k)^{-\frac{1}{1+p}}\left(\frac{1-(2k+1)q}{1-q}\right)\left(\frac{\text{Li}_{-p}(q)}{p!}\right)^{-\frac{1}{1+p}}.
    \label{}
\end{align}
It is straightforward to compute the free energy in this case, similar to the Jordan quiver.

\subsection{$\cN=4$ superconformal index}
In this part we consider the superconformal indices of the four-dimensional gauge theories and their matrix integral representation.
In general, the $\cN=1$  superconformal index is defined by
\bea
\mathcal{I}(\mu_i)= \Tr_{S^3}{(-1)^F e^{-\beta \delta} e^{-\mu_i \mathcal{M}_i}},
\eea
where the trace is over the Hilbert space of the theory quantized on $S^3$, $F$ is the fermion number, $\beta$ is the inverse temperature, $\delta$ is the Hamiltonian defined by $\delta= \frac{1}{2}\{\mathcal{Q}, \mathcal{Q}^\dagger\}$ with a supercharge $\mathcal{Q}$ and $\mathcal{M}_i$ are the global symmetry generators that are annihilated by the super charge with the associated fugacity $\mu_i$.

In particular, we consider $4d$ $\mathcal{N}=1$ superconformal field theory on $S^3\times S^1$. The global symmetry of the theory, i.e. the isometry of the $S^3$, is $\text{Spin}(4)= \mathrm{SU}(2)_1\times \mathrm{SU}(2)_2$ with indices $\alpha=\pm 1, \dot \alpha = \pm 1$. In $\mathcal{N}=1$ supersymmetry in four dimensions, we have four supercharges and their conjugates $\mathcal{Q}_\alpha, \mathcal{S}^\alpha= \mathcal{Q}^{\dagger\alpha}, \tilde{\mathcal{Q}}_{\dot\alpha},
\tilde{\mathcal{S}}^{\dot\alpha}=\tilde{
\mathcal{Q}}^{\dagger\dot\alpha}$. We choose a particular supercharge $\mathcal{Q}= \mathcal{Q}_{-}$, which satisfies the algebra $\{\mathcal{Q}, \mathcal{Q}^{\dagger}\}= \Delta-2j_1+\frac{3}{2} r$, with $\Delta$ is the conformal dimension, $j_1$ ($j_2$) is Cartan generator of $\mathrm{SU}(2)_1$ ($\mathrm{SU}(2)_2$), and $r$ is the $\mathrm{U}(1)_r$ R-charge. Then we can define the superconformal index, using some fugacity factors $p$ and $q$, as
\bea
\mathcal{I}(p,q) =\Tr{(-1)^F e^{-\beta \delta} p^{\frac{1}{3}(\Delta+j_1)+j_2} q^{\frac{1}{3}(\Delta+j_1)-j_2}}. 
\eea
Using the fact that the only states with  $\delta=0$ contribute to the index and thus the index is independent of $\beta$, we obtain
\bea
\mathcal{I}(p,q) =\Tr{(-1)^F p^{j_1+j_2-\frac{1}{2}r} q^{j_1-j_2-\frac{1}{2}r}}. 
\eea

The computations of the superconformal index has two parts, first is to compute the single letter index $i_k$ of each supermultiplet which is labeled here by $k$, and then using the plethystic exponential (PE), defined in \eqref{PE def}, and matrix integral \eqref{index PE}, to compute the full index, 
\bea
\label{}
\mathcal{I}(p,q, V) = \int_{\mathrm{U}(N)} \hspace{-1em} \dd{U} \prod_k \text{PE}\ [i_k(p,q, U,V)],
\eea
where $U$ denotes the gauge group element and $V$ denotes the flavour group element. 

In $\mathcal{N}=1$ SCFT, the single letter index is the sum of the vector multiplet index $i_V$ and chiral multiplet index $i_S$~\cite{dolan2009applications},
\bea
i=i_V+ i_S,
\eea
where
\bea
\label{index vector chiral}
i_V(p,q,U)= -(\frac{p}{1-p}+ \frac{q}{1-q})\, \chi_{\text{adj}}(U), \quad i_S(p,q,U,V)= \frac{(pq)^\frac{1}{3}\chi_{\bar{\mathcal{R}}}(U,V)-(pq)^\frac{2}{3}\chi_{\mathcal{R}}(U,V)}{(1-p)(1-q)},\nonumber\\
\eea
and the adjoint character $\chi_{\text{adj}}$, and the bifundamental character $\chi_{\mathcal{R}}$, are given by
\bea
\chi_{\text{adj}}(U^n)=   \tr U_a^n \tr U_a^{\dagger n}, \quad \chi_{\mathcal{R}_{a\bar{b}}}(U^n)= \tr U_a^n \tr U_b^{\dagger n}.
\eea
Next, we explain the matrix integral representation of the index in the explicit example of $\mathcal{N}=4$ SYM. In this case, using Eqs.\eqref{index PE} and \eqref{index vector chiral} the matrix integral representation of the superconformal index becomes
\bea
\label{N=4 index MI}
\mathcal{I}(p,q) = \int_{\mathrm{U}(N)} \hspace{-1em} \dd{U} \exp\left[\sum_{n=1}^\infty  a_n (p,q) \tr U^n \tr U^{-n}\right],
\eea
where
\bea
a_n(p,q) =\frac{i(p^n, q^n)}{n} , \quad i(p,q) = \frac{2pq -p-q + 3(pq)^{\frac{1}{3}
}- 3(pq)^{\frac{2}{3}}}{(1-p)(1-q)}.
\eea
Single letter index can be written as
\bea
i(p,q) = 1-\frac{(1-(pq)^{\frac{1}{3}})^3}{(1-p)(1-q)}.
\eea
In the large $N$ limit, the above matrix integral \eqref{N=4 index MI} becomes an infinite product,
\bea
\lim_{N\to \infty} \mathcal{I}(p,q) = \prod_{n=1}^\infty \frac{1}{1-i(p^n, q^n)}.
\eea
At weak coupling, we can effectively approximate the superconformal index \eqref{N=4 index MI} by the generalized GWW model
\bea
\label{}
\mathcal{I}(p,q) \approx
\mathcal{Z}(p,q) = \int_{\mathrm{U}(N)} \hspace{-1em} \dd{U} \exp\left[\sum_{n=1}^\infty t_n (p,q)\left(\tr U^n+ \tr U^{-n}\right)\right],
\eea
and in terms of generalized GWW couplings, we obtain
\bea
t_n=a_n(p,q)=\frac{i(p^n, q^n)}{n} , \quad t_n =\frac{1}{n}\left(1-\frac{(1-(pq)^{\frac{n}{3}})^3}{(1-p^n)(1-q^n)}\right).
\label{N=4 coupling}
\eea
In the rest of this section, we perform some numerical analysis of the index.
\subsubsection*{GWW model in $\cN=4$ SYM}
First we consider consider the truncated case $n=1$; the GWW model with the $\cN=4$ SYM coupling $t_1$ from Eq.\eqref{N=4 coupling}. Notice that this case is dominant in the phase structure of the model. In this case $\mathcal{Z}(p,q)$ reduces to a effective theory of GWW model,
\bea
\label{}
\mathcal{Z}(p,q) \sim \int_{\mathrm{U}(N)} \hspace{-1em} \dd{U} \exp\left[t_1 (p,q) \left(\tr U+ \tr U^{-1}\right)\right].
\eea
In unrefined case $p=q\equiv x$, we have
\bea
i(x)= 1-\frac{(1-x^{\frac{2}{3}})^3}{(1-x)^2}, \qquad t_n=\frac{1}{n}\left(1-\frac{(1-x^{\frac{2n}{3}})^3}{(1-x^n)^2}\right).
\eea
Thus, the critical parameters can be obtained as
\begin{align}
    \alpha_2 =t_1=  \frac{3 x^{\frac{2}{3}}+4x+2 x^{\frac{4}{3}}}{(1+x^{\frac{1}{3}}+x^{\frac{2}{3}})^2} 
    \, , \qquad
    \beta =2t_1= \frac{6 x^{\frac{2}{3}}+8x+4 x^{\frac{4}{3}}}{(1+x^{\frac{1}{3}}+x^{\frac{2}{3}})^2}
    \, , \qquad
    x_c=1,
    \label{}
\end{align}
where we used the observation that the critical point $t_1^*=i(x_c)=1$ implies $x_c=1$.
Then, the double-scaling parameter $s= \lambda_2 N^{\frac{2}{3}}$ can be computed,
\begin{align}
    \lambda_2 =  \frac{1-\beta}{\alpha_2^{1/3}}= \frac{1+2x^{\frac{1}{3}}-3x^{\frac{2}{3}}-6x-3x^{\frac{4}{3}}}{(1+x^{\frac{1}{3}}+x^{\frac{2}{3}})^{\frac{4}{3}}(3x^{\frac{2}{3}}+4x+2x^{\frac{4}{3}})^{\frac{1}{3}}}
       .
    \label{}
\end{align}
Using the parameter $\lambda_2$, the free energy in different regimes can be computed explicitly. However, as our result is comparable to the free energy obtained by other plausible methods in the vicinity of the critical point, it is natural to expand around the critical point $x_c=1$. Let us write the fugacity factor as $x=e^{\epsilon}$, in which the parameter $\epsilon$ is proportional to the radius of $S^1$ in the definition of the superconformal index, and it can be interpreted as the inverse temperature.
Thus,
we have the following high temperature expansions at small $\epsilon$,
\bea
t_1 = i(x) = 1+ \frac{8}{27}\epsilon - \frac{2}{243}\epsilon^3 + O(\epsilon^5),
\eea
and
\bea
\lambda_2= -1-\frac{40}{81}\epsilon+\frac{256}{6561}\epsilon^2+\frac{10606}{1594323}\epsilon^3 + O(\epsilon^4).
\eea
Using the above expansion of the parameters, the free energy can be computed from  Eq.\eqref{free energy gen},
\bea
\cF=\begin{cases}
    \displaystyle
        \cF_c+\cF_{\text{inst}} & \text{for } x<1\\[1em] \displaystyle
       \cF_c-\frac{1}{12} |\lambda_2|^3+\cF_{\text{pert}} & \text{for } x>1
        \end{cases},  
        \label{}
\eea
where we obtain
\bea
\cF_c= t_1^2 = 1+ \frac{16}{27}\epsilon +\frac{64}{729} \epsilon^2- \frac{4}{243} \epsilon^3 + O(\epsilon^4),
\eea
\bea
\cF_c-\frac{1}{12} |\lambda_2|^3= \frac{13}{12}+ \frac{58}{81}\epsilon +\frac{304 }{2187}\epsilon^2- \frac{2093 }{118098}\epsilon^3 + O(\epsilon^4),
\eea
and $\cF_{\text{inst}}$ and $\cF_{\text{pert}}$ can also be computed explicitly, as before.
\subsubsection*{Generalized GWW model in $\cN=4$ SYM}
Next, we consider the generalized GWW model approximation for the $\cN=4$ SYM.
Expanding around $x_c=1$, we similarly obtain
\bea
n\, t_n = i(x^n) = 1+ \frac{8n}{27}\epsilon - \frac{2n^3}{243}\epsilon^3 +O(\epsilon^5).
\eea
The expansion can be written formally as
\bea
n\, t_n = 1+ \sum_{l=1}^\infty (-1)^{l+1} a_l\ \epsilon^{2l-1} n^{2l-1},
\eea
for some positive numerical coefficients $a_l$. Thus, the parameters $\alpha_p$ and $\beta$ defined in Eq.\eqref{alpha beta p}, can be evaluated as
\bea
\beta&=& 2\sum_{n=1}^{\infty}\left(1+ \sum_{l=1}^\infty (-1)^{l+1} a_l\ \epsilon^{2l-1} n^{2l-1}\right)=2\zeta(0) +2 \sum_{l=1}^\infty (-1)^{l+1} a_l \epsilon^{2l-1} \zeta(1-2l),\nonumber\\
\alpha_p &=& \frac{2}{p!}\sum_{n=1}^{\infty}\left(n^p+ \sum_{l=1}^\infty(-1)^{l+1}a_l\ \epsilon^{2l-1} n^{p+2l-1}\right)= \frac{2}{p!} \zeta(-p) + \frac{2}{p!} \sum_{l=1}^\infty (-1)^{l+1} a_l \epsilon^{2l-1} \zeta(1-p-2l)
.\nonumber\\
\eea
To be explicit, we keep the few terms in the expansion as following
\bea
\beta&=& 2\zeta(0)+\frac{16}{27}\zeta(-1)\epsilon-\frac{4}{243}\zeta(-3) \epsilon^3,\nonumber\\
\alpha_p &=& \frac{2}{p!}\zeta(-p)+ \frac{16}{27p!}\zeta(-1-p)\epsilon-\frac{4}{243p!}\zeta(-3-p) \epsilon^3,
\eea
and thus the parameter $\lambda_p$ can be obtained from $\lambda_p =\alpha_p^{-\frac{1}{1+p}}(1-\beta)$.
The bulk free energy $\cF_c$ can be computed similarly,
\bea
\cF_c= \sum_{n} n\, t_n^2 = \sum_{l=1}^{\infty} b_l \zeta(-l) \epsilon^{l-1},
\eea
and the numerical coefficients can be computed explicitly 
\bea
b_1=1,\quad b_2=16/27,\quad b_3=64/729,\quad b_4=-4/243, \quad \ldots.
\eea
One can use this data to compute explicitly the free energy in different phases. In $p=2$ case, up to order three, we have
\bea
\cF=\begin{cases}
    \displaystyle
        \cF_c & \text{for } x<1\\[1em] \displaystyle
       \cF_c-\frac{1}{12} |\lambda_2|^3 & \text{for } x>1
        \end{cases},  
        \label{}
\eea
where 
\bea
\cF_c = -\frac{1}{12}+\frac{8}{10935} \epsilon^2 + O(\epsilon^4),
\eea
\bea
\cF_c-\frac{1}{12} |\lambda_2|^3= -\frac{241}{12}-\frac{270}{\epsilon}+ \frac{1745}{567}\epsilon +\frac{283363 }{1377810}\epsilon^2- \frac{9313 }{214326}\epsilon^3 + O(\epsilon^4).
\eea




\section{Discussion and future research}
In this work, we discuss some universal results for the generic matrix models, including the double-trace matrix model, up to some model dependent parameters, and we compute these parameters explicitly for the generalized GWW model.
The main universal results of this study in the finite but large $N$ and infinite limit of free $\mathrm{U}(N)$ gauge theory partition functions and the indices of gauge theories include:
\begin{itemize}

\item Universal critical/multi-critical, large and finite $N$ results such as $1/N$ expansion, genus expansion, instanton sector, etc. in generic unitary matrix model.

\item  Universality in the deconfinement and Hagedorn phase transition in the infinite $N$ limit of generic unitary matrix models.

\item Universal multi-critical edge fluctuation and  phase transitions in generic unitary matrix models. Infinite $p$ limit of the multi-critical dynamics and the emergence of bulk fluctuation. The
interpretation of the results in terms of the gap dynamics.
\end{itemize}

In this work, we studied the matrix models with real couplings. Moreover, the general case of the complex coupling is of great interest. This line of research has recently  attracted a lot of interest in the theoretical physics community because of its relations to black holes in the context of AdS/CFT, for a review see \cite{zaffaroni2020ads} and references therein. 

In this work, we considered one-matrix models and corresponding quiver gauge theories with one node, however, the general case of the quiver gauge theories deal with the multi-matrix model.
A possible future extension of this work would be to consider the superconformal index of quiver gauge theories and their multi-matrix integral representation.
In particular, the two-matrix model for conifold quiver and related gauge theories would be the first step in this direction of research.


In the study of the right tail of the TW distribution in the instanton sector, we used the Airy function approximation to expand the Fredholm determinant. An alternative approach would be to consider the approximation based on the Bessel function and its variants. The possible result in this way, would be of great interest because of its comparison with the existing results about the instantons of the unitary matrix models in the literature.

In the multi-critical dynamics, the asymptotic analysis lacks the rigorous results for the sub-leading corrections to the Fredholm determinant of the higher Airy kernel. Developing techniques and results in this direction are highly interesting.

In this work, the main focus of the asymptotic analysis is the computation of the free energy, however, an equally important parallel direction is the computations of the limit shape and the entropy using the spectral curve method and the large deviation technique.

The results of this study is based on the analysis for the case of ($p\in 2\mathbb{N}$)-multi-critical dynamics. An immediate continuation of this analysis is to consider the odd $p$ case and its applications and implications for random partitions and gauge theories.

In this work, we studied the unitary matrix model and their phase structure. A natural continuation of this work is to consider Hermitian matrix model. The multi-critical analysis for the Hermitian matrix model is performed in \cite{Claeys:2009CPAM}. Possible applications of this result are in the weakly coupled gauge theories on compact manifolds \cite{Aharony:2003sx} and their phase structure. Another interesting model is the Douglas--Kazakov matrix model with quadratic potential and its instanton sector. A straightforward generalization of our results would shed light to this classic matrix model.
In this work we considered the same coefficients for the GWW and the double-trace matrix model, which is plausible for the effective theory at weak coupling. However, one can study the relations between the GWW and double-trace model and their phase transition using the Legendre transform and saddle-point analysis. 
One can use the Hubbard--Stratonovich transformation and apply our results for the generalized GWW model to the double-trace matrix integrals and their gravity duals. Asymptotic analysis of the double-trace matrix models with real and complex couplings is an interesting direction and remains for future. In this way, based on the rigorous mathematical results, we hope to provide new understandings for the phase structure associated to the double-trace models such as deconfinement, Hagedorn and Hawking--Page phase transitions \cite{Copetti:2020dil} and their multi-critical generalizations.
Especially, possible interpretations of the multi-critical phase structure for the Hagedorn and Hawking-Page transitions would be interesting.

In this regard, the GWW model and its generalization seems to explain the transition between highly excited string states and black holes, whereas the double-trace model explains the Hawking-Page transition between the thermal AdS and black holes \cite{alvarez2006black}. Thus, our results about the critical dynamics of the generalized GWW model studied in this paper perhaps directly explore some universality features of the former transition and deserves further studies. However, the highly interesting question of the possible interpretations and implications of the multi-critical dynamics in the gravity side is not clear at this moment, which would require new ideas and techniques. Regarding the Hawking-Page transition, we should apply the Hubbard-Stratonovich transformation to discuss all possible implications of the critical and multi-critical dynamics of the double-trace model in the gravity side. We will return to this issue in a future work.

\subsubsection*{Acknowledgments}

This work has been supported in part by ``Investissements d'Avenir'' program, Project ISITE-BFC (No.~ANR-15-IDEX-0003), EIPHI Graduate School (No.~ANR-17-EURE-0002), and Bourgogne-Franche-Comté region.
\bibliography{ref}

\end{document}